\documentclass[sigconf]{acmart}

\AtBeginDocument{%
  \providecommand\BibTeX{{%
    \normalfont B\kern-0.5em{\scshape i\kern-0.25em b}\kern-0.8em\TeX}}}

\copyrightyear{2025}
\acmYear{2025}
\setcopyright{cc}
\setcctype{by}
\acmConference[CHI '25]{CHI Conference on Human Factors in Computing Systems}{April 26-May 1, 2025}{Yokohama, Japan}
\acmBooktitle{CHI Conference on Human Factors in Computing Systems (CHI '25), April 26-May 1, 2025, Yokohama, Japan}
\acmDOI{10.1145/3706598.3713218}
\acmISBN{979-8-4007-1394-1/25/04}

\usepackage{framed}
\usepackage{tabularx}
\usepackage{graphicx}
\usepackage{multirow}
\usepackage{color, colortbl}
\usepackage{subfigure}
\usepackage{listings}
\usepackage{parcolumns}

\newcommand{\glcomment}[1]{}

\newcommand{\ujcomment}[1]{}

\newcommand{\gdcomment}[1]{}
\newcommand{\revise}[1]{#1}

\newcommand{\ignore}[1]{}

\usepackage{xspace}
\newcommand{\etal}{\emph{et al.}\xspace}
\newcommand{\paratitle}[1]{\vspace{1.0ex}\noindent\textbf{#1}}
\newcommand{\ie}{\textit{i.e.,}~}
\newcommand{\eg}{\textit{e.g.,}~}

\begin{document}

\title{Plan-Then-Execute: An Empirical Study of User Trust and Team Performance When Using LLM Agents As A Daily Assistant}

\author{Gaole He}
\affiliation{%
  \institution{Delft University of Technology}
  \city{Delft}
  \country{The Netherlands}}
\email{g.he@tudelft.nl}

\author{Gianluca Demartini}
\affiliation{%
  \institution{The University of Queensland}
  \city{Brisbane}
  \country{Australia}
}
\email{g.demartini@uq.edu.au}

\author{Ujwal Gadiraju}
\affiliation{%
  \institution{Delft University of Technology}
  \city{Delft}
  \country{The Netherlands}
}
\email{u.k.gadiraju@tudelft.nl}
\renewcommand{\shorttitle}{An Empirical Study of User Trust and Team Performance with LLM Agents As A Daily Assistant}

\begin{abstract}
Since the explosion in popularity of ChatGPT, large language models (LLMs) have continued to impact our everyday lives. 
Equipped with external tools that are designed for a specific purpose \revise{(\eg for flight booking or an alarm clock)}, LLM agents exercise an increasing capability to assist humans in their daily work. 
Although LLM agents have shown a promising blueprint as daily assistants, there is a limited understanding of {how they can provide daily assistance based on planning and sequential decision making capabilities}. We draw inspiration from recent work that has highlighted the value of `\textit{LLM-modulo}' setups in conjunction with humans-in-the-loop for planning tasks.
We conducted an empirical study ($N$ = 248) of LLM agents as daily assistants in six commonly occurring tasks with different levels of risk typically associated with them (\eg flight ticket booking and credit card payments). 
To ensure user agency and control over the LLM agent, we adopted LLM agents in a plan-then-execute manner, wherein the agents conducted step-wise planning and step-by-step execution \revise{in a simulation environment}. 
We analyzed how user involvement at each stage affects their trust and collaborative team performance. 
Our findings demonstrate that LLM agents can be a double-edged sword --- (1) they can work well when a high-quality plan and necessary user involvement in execution are available, and (2) users can easily mistrust the LLM agents with plans that seem plausible. 
We synthesized key insights for using LLM agents as daily assistants to calibrate user trust and achieve better overall task outcomes. 
Our work has important implications for the future design of daily assistants and human-AI collaboration with LLM agents.
\end{abstract}




\begin{teaserfigure}
    \centering
  \includegraphics[width=0.75\textwidth]{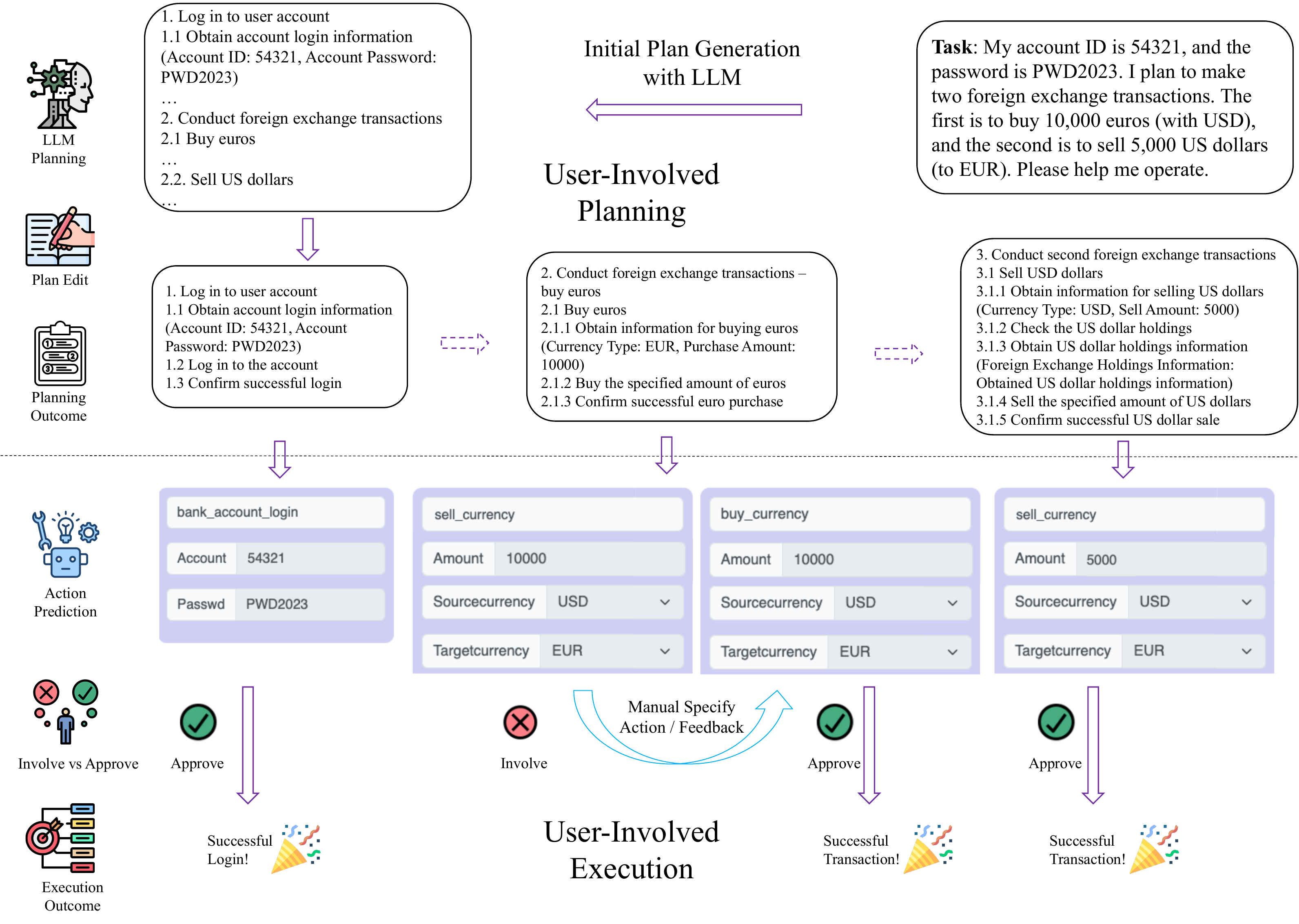}
  \caption{Illustration of the human-AI collaboration with plan-then-execute LLM agents.}
  \Description{Illustration of the human-AI collaboration with plan-then-execute LLM agents. The planning stage mainly consists of two steps: (1) LLMs generate a draft plan (2) user edit the planning. Based on the edited plan, we get a step-wise plan (three steps in illustration). Then, in the execution stage, LLM agents will generate a sequence of actions, which is one-on-one mapping based on the plan's primary steps. Users will join this process to decide whether they approve the predicted action or choose to be involved (with manual specification of action or give text feedback to LLM agents). After the step-by-step execution of the plan, the task is solved.}
  \label{fig:illustration}
\end{teaserfigure}

\maketitle

\section{Introduction}

Autonomous agents have been regarded as a research focus for artificial intelligence (AI) over the last century~\cite{albrecht2018autonomous}. 
With the wish that autonomous agents can make our life better, many autonomous agents have been designed as virtual personal assistants~\cite{kepuska2018next}. 
These AI assistants (\eg Siri) perform well (albeit imperfectly) in following user instructions to execute low-risk tasks like playing a song, reporting weather forecasts, or searching for an image to support everyday tasks. 
However, on tasks entailing potential risks (\eg monetary payments or hiring an employee), humans hesitate to trust such AI systems due to loss aversion~\cite{tversky1991loss} and algorithmic aversion~\cite{hou2021expert,mahmud2022influences,dietvorst2015algorithm}. 
Only when users can obtain a sense of control by being able to modify the outcomes of imperfect AI can they overcome such algorithm aversion and be willing to collaborate with imperfect AI systems~\cite{dietvorst2018overcoming}. 

With the recent rise of large language models (LLMs) in natural language understanding and generation~\cite{zhao2023survey}, researchers have started to analyze LLM-based agents and their applicability in a plethora of tasks 
~\cite{xi2023rise,wang2024survey}. 
The term `\textit{LLM agent}' refers to an artificial entity based on LLMs that perceives its context, makes decisions, and then takes actions in response~\cite{xi2023rise}.
Compared to existing deep learning and LLM-based methods (\eg chaining multiple LLMs~\cite{wu2022ai}), LLM agents provide more flexibility in task solving and user interaction, which makes them suitable for daily assistance. This is primarily due to three reasons.
First, with a planning module, LLM agents can generate a dynamic plan based on the tools provided~\cite{xi2023rise,wang2024survey}. 
Such plans are typically defined in a logical structure --- step-wise plans, which can be easily understood by humans. 
Second, with LLMs as a core control module, users can access and interact with external toolkits via a more natural interaction (\ie conversation) with LLM agents~\cite{bommasani2021opportunities,zhao2023survey}, reducing manual control efforts over function-specific tools. 
For example, LLM agents can complete time-consuming jobs like information seeking and information filtering (\eg searching for a flight in itinerary planning) based on specific user needs. 
Third, the Markov decision process of LLM agents can generate a sequence of actions (\ie using external toolkits)\footnote{In our study, the usage of one tool is the same as executing one action. Therefore, we refer to a tool and action interchangeably.} as output. 
Paired with an understanding of actions and necessary parameters for the interaction with the LLM agents, users can get involved in the real-time execution of tasks with LLM agents and fix potential problems while benefiting from task delegation~\cite{lubars2019ask}. {Based on an intuitive framework for task delegation, Lubars \etal~\cite{lubars2019ask} found that user trust can play an important role in human delegation behaviors to AI systems. However, there is a relatively limited understanding of user trust development and calibration in collaboration with LLM agents.} 




There is also a growing debate in the machine learning and AI research communities about whether LLMs can be truly considered as reasoning and planning agents~\cite{kambhampati2024llms}. With this in the backdrop, existing work on automated task completion has revealed that LLM agents can exhibit promising performance in handling complex tasks like playing games~\cite{wang2023describe}, answering complex questions~\cite{zhuang2024toolqa}, and in simulating social behavior~\cite{park2023generative}. 
However, such agents are still far from perfect. Due to the probabilistic nature of LLMs, there is much uncertainty in automating LLM agents for tasks with high risks attached. 
To avoid unintended or unexpected consequences, 
there is a need for user control over the real-time execution process. 
{Through an empirical study of LLM planning capabilities, planning experts found that ``LLMs’ ability to generate executable plans autonomously is rather limited''~\cite{valmeekam2023planning}. 
However, when combined with a sound planner in an `LLM-Modulo' mode, ``the LLM-generated plans can improve the search process for underlying sound planners''~\cite{valmeekam2023planning}. 
Humans can potentially be the `sound planners' who can work in conjunction and optimize plans drafted by LLMs, which can then be executed by LLM agents. 
Such human-AI collaboration can reduce human efforts in generating a reliable plan from scratch. 
} 

Attracted by the promise of LLM agents, there have been some early explorations~\cite{geissler2024concept} of adopting them in human-AI collaboration. 
However, the existing works have mainly analyzed how LLM agents can serve specific use cases (\eg design creation~\cite{geissler2024concept}), and others have conducted structured interviews to obtain expert insights~\cite{zhang2024s,zheng2023synergizing}. 
It is still unknown how well LLM agents can work as general daily assistants to assist users in everyday tasks with varying stakes and how user trust and team performance are shaped by interacting with LLM agents. 

In our work, we address this research gap and adopt LLM agents to assist humans in everyday tasks by following a 
plan-then-execute workflow~\cite{wang2023plan}. 
First, the LLM agent generates a step-wise plan formulated with a hierarchical structure. 
Then, the LLM agent executes the generated plan by transforming it into a sequence of actions (leveraging external toolkits). 
The benefits of such a plan-then-execute framing are three-fold: 
(1) Compared to a dynamic process where planning and execution are bound closely, separating planning and execution into two stages provides more task clarity to the users, which reduces user cognitive load and contributes to the quality of task outcomes~\cite{gadiraju2017clarity}. 
(2) With planning at the beginning of the task, users can develop a global understanding of how the LLM agents will execute the task. 
Based on a follow-up step-by-step execution, it would be straightforward for users to be involved in such a process and control the outcomes of task execution. 
(3) Planning and execution are representative abstractions of how LLM agents work. 
The findings of such an empirical study can be generalized to human-AI collaboration with other kinds of LLM agents (\eg dynamic planning-execution). 
%
To this end, we propose the following research questions:
\begin{itemize}
\item \textbf{RQ1}: How does human involvement in the high-level planning and real-time execution shape their trust in an AI system powered by LLM agents?
\item \textbf{RQ2}: How does human involvement in the high-level planning and real-time execution of tasks with an AI system powered by LLM agents affect the overall task performance?
\end{itemize}

Addressing these research questions, we carried out an empirical
study ($N$ = 248) of human-AI collaboration in six different everyday scenarios with varying stakes and risks attached (\eg credit card payment and itinerary planning). 
We found that user involvement in the planning and execution can be beneficial in addressing imperfect plans and fixing execution errors. 
As a result, LLM agents can achieve better task performance. 
However, we also found that user involvement in the planning and execution stages of the LLM agent fails to calibrate user trust in corresponding task outcomes. 
{A potential reason here is that the plausible plans generated by the LLMs can mislead users into trusting the LLM agents when they are in fact wrong.} 
Our findings highlight that user involvement can also bring about additional trade-offs to consider: 
(1) user involvement in the planning and execution poses a high cognitive load on users and decreases user confidence in their decisions; (2) user involvement can be harmful in some task contexts (\eg user involvement reduces plan quality). 
Further research is required to understand when to provide necessary user involvement.
%
Our key insight is that as opposed to following a fixed mode of user involvement, it is prudent to explore how user involvement in planning and execution can be tailored to fit the task and the user. 
Based on our quantitative and qualitative findings, 
we share insights for designing effective LLM agents as daily assistants and synthesize promising directions for further research around LLM agents in the context of human-AI collaboration. 
Our work has important theoretical implications for human-AI collaboration with LLM assistance and design implications for plan-then-execute LLM agents to support human-AI collaboration.
\section{Background and Related Work}
Our work proposes to analyze how user involvement in the planning and execution stages of LLM agents shapes user trust in the LLM agents and the overall task performance of LLM agents. 
Thus, we position our work in \revise{three realms} of related literature: human-AI collaboration (\S~\ref{sec-rel-collaboration-LLM}), \revise{trust and reliance on AI systems (\S~\ref{sec-rel-trust-reliance}),} task support with LLMs and LLM agents (\S~\ref{sec-rel-LLM-agent}). 


\subsection{Human-AI Collaboration}
\label{sec-rel-collaboration-LLM}

In recent decades, deep learning-based AI systems have shown promising performance across various domains~\cite{yang2022survey,fernando2021deep} and applications~\cite{pouyanfar2018survey,dong2021survey}. 
However, such AI systems are not good at dealing with out-of-distribution data~\cite{jia2017adversarial,mccoy-etal-2019-right}, and their intrinsic probabilistic nature brings much uncertainty in 
practice~\cite{ghahramani2015probabilistic}. 
Such observations raise wide concerns about the accountability and reliability of AI systems~\cite{kaur2022trustworthy}. 
Under such circumstances, human-AI collaboration has been recognized as a well-suited approach 
to taking advantage of their promising predictive power and ensuring trustworthy outcomes~\cite{lai2021towards,jiang2021supporting}. 
While humans can provide more reliable and accountable task outcomes, too much user involvement to check and control AI outcomes is undesirable~\cite{lai2022human}. 
It goes against the premise that AI systems are introduced to reduce human workload. 
In that context, researchers have theorized and empirically analyzed when and where users could and should delegate to AI systems~\cite{lai2022human,lubars2019ask}. 

\paratitle{Task Delegation}. While humans prefer to play the leading role in human-AI collaboration~\cite{lubars2019ask}, delegating to AI systems can bring benefits like cost-saving and higher efficiency. 
Apart from manual delegation decisions, it is common to apply automatic rules for human delegation (\eg heuristics obtained from domain expertise or manually crafted rules~\cite{lai2022human}).
Many user factors like trust~\cite{lubars2019ask}, human expertise domain~\cite{erlei2024understanding}, and AI knowledge~\cite{pinski2023ai}) have a substantial impact on human delegation behaviors. 
Another relevant stream of recent research has explored AI delegation to humans~\cite{madras2018predict,fugener2022cognitive,pinski2023ai}. 
Researchers have investigated the conditions under which AI systems should defer to a human decision maker, which may bring benefits of improved fairness~\cite{madras2018predict}, accuracy~\cite{narasimhan2022post}, and complementary teaming~\cite{ijcai2022p344}. 
Compared to human delegation, AI delegation has been observed to achieve more consistent benefits in team performance~\cite{fugener2022cognitive,hemmer2023human}. {In collaboration with LLM agents, users need to determine when they should be involved in high-level planning and real-time execution. Such involvement decisions are similar to the delegation choices made by users. While task delegation is not the focus of our study, future work can explore this further.}

\paratitle{AI-assisted Decision Making} has attracted a lot of research focus in human-AI collaboration literature. 
Most existing work has conducted empirical studies~\cite{lai2021towards} and structured interviews~\cite{jiang2021supporting} to understand how factors surrounding the user, task, and AI systems affect human-AI collaboration. 
User factors like AI literacy~\cite{Chiang-IUI-2022}, cognitive bias~\cite{rastogi2022deciding}, and risk perception~\cite{fogliato2021impact,green2021algorithmic} have been observed to substantially impact user trust and reliance on the AI system. 
Similarly, task characteristics like task complexity and uncertainty~\cite{salimzadeh2023missing,salimzadeh2024dealing} and factors of the AI system (\eg performance feedback~\cite{bansal2019beyond,Lu-CHI-2021}, AI transparency~\cite{vossing2022designing} and confidence of AI advice~\cite{tomsett2020rapid,Zhang-FAT-2020}) also affect user trust and reliance on the AI system. 
For a more comprehensive survey of existing work on AI-assisted decision making, readers can refer to~\cite{lai2021towards}.

While machine learning and deep learning methods have been extensively analyzed in existing human-AI collaboration literature, to our knowledge, human-AI collaboration with LLM agents is still under-explored. 
Unlike previous studies where AI systems only follow a fixed mode to generate advice, LLM agents can be equipped with more logical clarity and can provide a step-wise plan and can follow a step-by-step execution. 
With such a plan-then-execute setup, LLM agents can bring high flexibility as well as uncertainty in high-level planning and real-time execution. Little is known about
how well LLM agents can work as daily assistants while handling tasks entailing varying stakes and potential risks. 
In our study, we analyzed the impact of user involvement in such AI systems by adjusting their intermediate outcomes (plan and step-by-step execution) to calibrate their trust and improve task outcomes. 
Our findings and implications can help advance the understanding of the effectiveness of LLM agents in human-AI collaboration.


\subsection{\revise{Trust and Reliance on AI systems}}
\label{sec-rel-trust-reliance}
\revise{Trust and reliance have been important research topics since human adoption of automation systems~\cite{lee2004trust,dzindolet2003role}. Due to the widespread integration of AI systems in nearly all walks of society, 
there has been a growing interest in understanding user 
trust~\cite{vereschak2021evaluate,mehrotra2024systematic} and reliance~\cite{eckhardt2024survey} on AI systems.}
User trust in the context of human-AI collaboration is typically operationalized as a subjective attitude toward AI systems/AI advice~\cite{lee2004trust}. 
In comparison, user reliance on AI systems is based on user behaviors (\eg adoption of AI advice and modification of AI outcomes). 
\revise{The two constructs have been shown to be highly related~\cite{lee1992trust,lee2004trust}: for example, user trust can substantially affect user reliance~\cite{lee2004trust}. 
However, they are intrinsically different and cannot be viewed as a direct reflection of each other~\cite{kahr2024understanding}. 
Most existing work has, therefore, studied the two constructs separately in terms of subjective trust and objective reliance.}

\revise{
Earlier work exploring human-AI trust primarily focused on the impact of different contextual factors surrounding user (\eg risk perception~\cite{green2021algorithmic}), task (\eg task complexity~\cite{salimzadeh2023missing}), and system (\eg stated accuracy~\cite{yin2019understanding,Zhang-FAT-2020}). 
Empirical studies have shown that most users tend to trust AI systems that are perceived to be highly accurate~\cite{yin2019understanding}. 
Such trust is vulnerable, as the AI system may provide an illusion of competence with persuasive technology (\eg explanations~\cite{chromik2021think,He-IUI-2025}) or overclaimed performance~\cite{yin2019understanding}. 
Even if the AI systems are accurate on specific datasets, they still suffer from out-of-distribution data~\cite{liu2021understanding,chiang2021you}. 
The misplaced trust in the AI systems can lead to misuse of the systems.
Several empirical studies~\cite{tolmeijer2021second} have shown that once users realize the AI system errs or performs worse than expected, their trust in the AI system can be violated, 
even resulting in the disuse of the AI system. 
Both the misuse and disuse of the AI system hinder optimal human-AI collaboration. 
}

\revise{
To address such concerns, researchers have explored how to help users calibrate their trust in the AI system. 
Different techniques to help users realize the trustworthiness of the AI system have been proposed~\cite{kaur2022trustworthy,rechkemmer2022confidence,ma2023should}. 
For example, increasing the transparency of AI systems by providing confidence scores~\cite{Zhang-FAT-2020}, explanations~\cite{wang2021explanations}, trustworthiness cues~\cite{liao2022designing}, and uncertainty communication~\cite{Sunnie-FAccT-2024}. 
However, the actual trustworthiness of the AI system does not always align with user perception. 
As found by \citet{banovic2023being}, untrustworthy AI systems can deceive end users to gain their trust. 
Another example is that users can develop an illusion of explanatory depth brought by explainable AI techniques~\cite{chromik2021think}, which leads to uncalibrated trust in the AI system. 
Even if users have indicated trust in the AI system, they may turn to rely more on themselves in final decision-making. 
The reasons are complex, and many factors, such as accountability concerns~\cite{lima2021human,tolmeijer2022capable} and cognitive bias~\cite{he2023knowing}, may affect user reliance behaviors. 
}

While trust calibration is an important goal in human-AI collaboration, it may be not enough to ensure complementary team performance. 
Through empirical user studies with different confidence levels of AI predictions, Zhang \etal~\cite{Zhang-FAT-2020} found that ``trust calibration alone is not sufficient to improve AI-assisted decision making''. 
To achieve optimal human-AI collaboration, humans and AI systems need to play complementary roles~\cite{hemmer2021human,hemmer2024complementarity}, and humans need to know when they should adopt AI assistance. 
In other words, humans should rely on AI advice when AI systems are correct and outperform them, and override AI advice when AI systems are incorrect or less capable than humans. 
Such user reliance patterns are denoted as \textit{appropriate reliance}~\cite{schemmer2022should,schemmer2023appropriate}, which is the key to
achieving complementary team performance. 

The main issues that lead to sub-optimal human-AI collaboration are: under-reliance (\ie disuse AI assistance when AI systems outperform humans) and over-reliance (\ie misuse AI assistance when AI systems are wrong or perform worse than humans)~\cite{schemmer2022should}. 
Users with an uncalibrated trust in the AI system can be easily misled to disuse or misuse AI systems~\cite{jacovi2021formalizing}. 
Researchers have proposed various interventions to promote appropriate reliance~\cite{he2023knowing,Lu-CHI-2021,lu2024does,chiang2021you,Chiang-IUI-2022} and calibrate user trust in AI systems~\cite{buccinca2021trust,Zhang-FAT-2020}.  
For example, explainable AI methods have been shown to help reduce over-reliance~\cite{vasconcelos2023explanations} and under-reliance~\cite{wang2021explanations} in different scenarios albeit with little consistency across contexts. 
Another example is tutorial interventions, which have shown effectiveness in user onboarding~\cite{lai2020chicago}, mitigating cognitive biases~\cite{he2023knowing} and developing AI literacy~\cite{Chiang-IUI-2022}. 
For a more comprehensive overview of interventions to facilitate trust calibration and appropriate reliance, readers can refer to ~\cite{lai2021towards,eckhardt2024survey,mehrotra2024systematic,kahr2024understanding}.

\revise{LLM agents~\cite{wang2024survey} have gained much popularity in recent years, distinguishing them from most prior AI systems. 
They can communicate through conversation, plan logically, and can be built to leverage powerful external tools to achieve complex functions.
While trust and reliance have been extensively analyzed in existing human-AI collaboration literature, it is still unclear how users trust and rely on AI systems powered by LLM agents. 
In our work, calibrated trust is adopted as an important goal for human-AI collaboration in the planning and execution stage. 
Meanwhile, users are expected to fix potential errors in the planning and execution stages, reflecting their reliance on the AI system. 
Our work can substantially advance the understanding of trust and reliance on plan-then-execute LLM agents.
}


\subsection{Task Support with LLMs and LLM Agents}
\label{sec-rel-LLM-agent}
LLMs and LLM agents bring new opportunities and challenges to human-AI collaboration~\cite{bommasani2021opportunities}. 
It is evident that their generation capabilities can help reduce the cognitive effort from humans. 
But LLMs are also riddled with challenges such as hallucination~\cite{ji2023survey} (\ie generated text seems plausible but is factually incorrect). 
Failure to handle such issues may bring fatal errors with unaffordable costs depending on the context (\eg medical diagnosis). 

\begin{figure*}[h]
    \centering
    \includegraphics[width=0.75\textwidth]{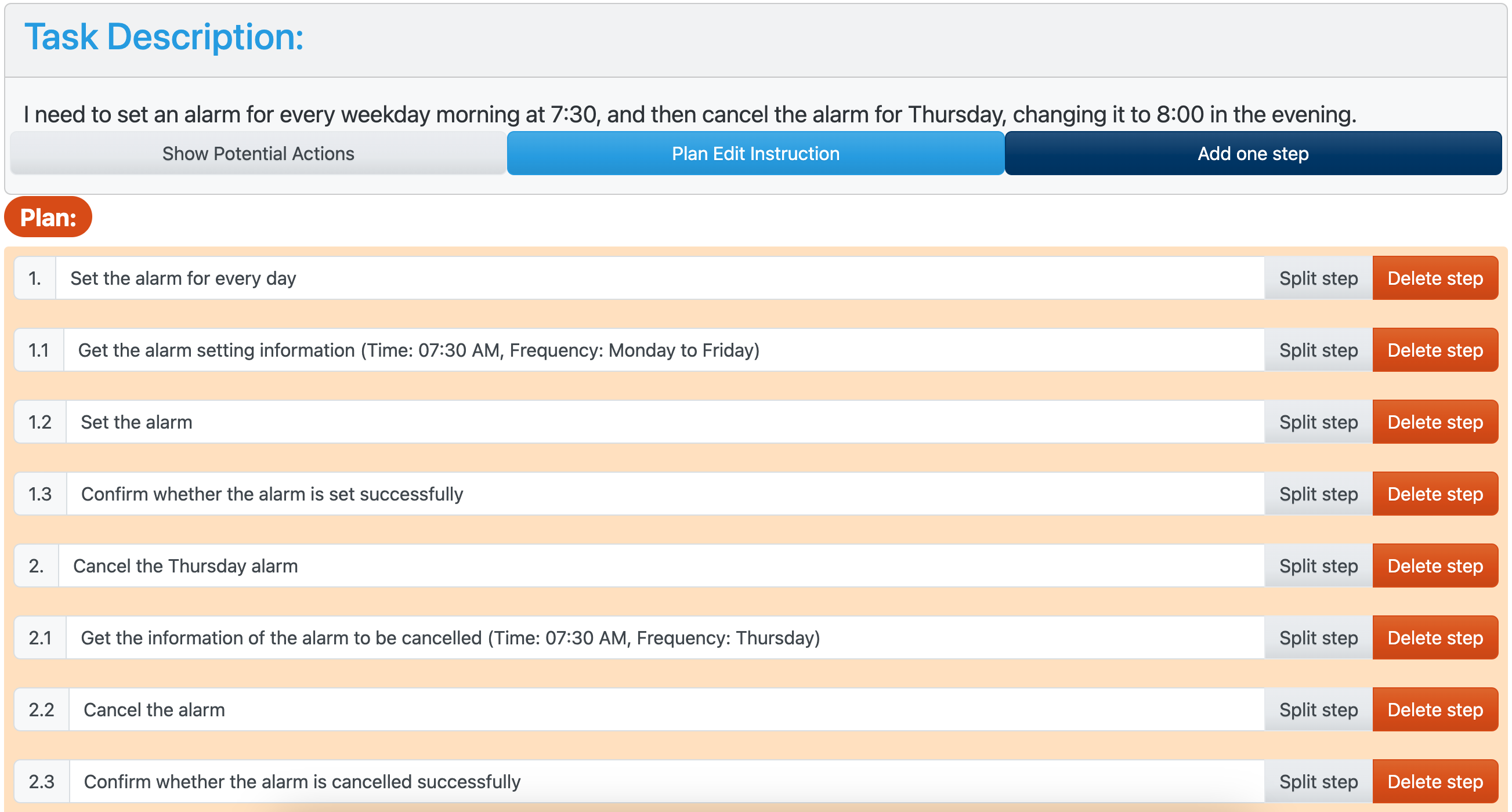}
    \caption{Screenshot of user-involved planning interface.} 
    \label{fig:planning}
    \Description{Screenshot of user-involved planning interface. At the top, we show the task description along with three buttons: show potential actions, plan edit instruction, and add one step. At the bottom, we show a step-wise plan for setting alarms. Users can click these buttons to achive the function we described in the user-involved planning to edit the plan.}
\end{figure*}

Due to the capability of generating coherent, knowledgeable, and high-quality responses to diverse human input~\cite{wei2022emergent}, a wide community of human-computer interaction researchers has paid attention to large language models~\cite{liao2023ai}.
Researchers have actively explored how LLMs can assist users in various tasks like data annotation~\cite{wang2024human,he2024if}, programming~\cite{omidvar2024evaluating}, scientific writing~\cite{shen2023convxai}, and fact verification~\cite{si2024large}. 
All the above functions can be achieved with elaborate prompt engineering using a single LLM. 
By chaining multiple LLMs with different functions, humans can customize task-specific workflows to solve complex tasks~\cite{wu2022ai}. 
Apart from obtaining answers with a one-shot text generation, LLMs also provide convenient conversational interactions. 
Through empirical studies, such conversational interactions have been shown to be effective in human-AI collaboration with multiple applications, such as decision making~\cite{slack2023explaining,lin2024decision,ma2024towards}, scientific writing~\cite{shen2023convxai}, and mental health support~\cite{sharma2023human}. 
With the growing popularity of LLMs, more and more humans have begun to adopt LLMs (\eg ChatGPT) to boost their work efficiency and productivity 
~\cite{zhao2023survey}.


LLM agents have been shown to have good planning, memory, and toolkit usage capabilities~\cite{xi2023rise, wang2024survey}. 
When suitable toolkits are provided, LLM agents can readily generate a task-specific plan and solve the tasks using toolkits. 
Attracted by the promise of LLM agents, there have been some early explorations~\cite{geissler2024concept,zheng2023synergizing,zhang2024s} of adopting them in human-AI collaboration contexts. 
These works were mostly analyzed in specific use cases (\eg design creation~\cite{geissler2024concept}). 
It is unclear how user trust and team performance are affected by user interactions with LLM agents in a sequential decision making setup (\ie solving a task by executing a sequence of actions) where users can be in control of the execution. 
To fill this research gap and advance our understanding of user control over LLM agents, we carried out a quantitative empirical study. 




\section{Method}


\subsection{Overview of User Involvement in Plan-then-execute LLM Agents}

In our study, we adopted plan-then-execute LLM Agents~\cite{wang2023plan} as assistants to help users handle daily tasks, \eg itinerary planning and currency transactions. Figure~\ref{fig:illustration} illustrates how users collaborate with plan-then-execute LLM agents. 
First, the LLM agents will generate a step-wise plan based on a prompt specifying the plan format adopted from~\cite{huang2024planning}. 
Then, users will make necessary edits to the plan based on the provided edit tools (will be further detailed in Section~\ref{sec-method-planning}). 
After the user edit, we obtained the step-wise plan as outcomes of the planning stage. 
Next, the LLM agents will transform the step-wise plan into a sequence of action predictions, which will be served in a step-by-step manner. 
Users will join the real-time execution process and check whether they approve the current predicted action (\ie blue card shown in Figure~\ref{fig:illustration}) or they would like to modify the current action prediction. The user involvement in execution stages will be introduced in Section~\ref{sec-method-execution}. 
After the iterative execution of all steps, the task is solved. {The evaluation of task performance is mainly based on the plan quality and execution accuracy of the action sequences.}

\paratitle{Implementation details}. In our study, we adopted GPT-3.5-turbo as the backbone LLM to serve the plan-then-execute LLM agent. The backend LLM agent implementation is mainly based on the Langchain plan and execute agent.\footnote{\url{https://api.python.langchain.com/en/latest/plan_and_execute/langchain_experimental.plan_and_execute.agent_executor.PlanAndExecute.html}} \revise{The execution of tasks are based on a simulation environment, where all tools/actions of the LLM agents are pre-defined as backend APIs hosted with Flask\footnote{\url{https://github.com/pallets/flask}}.}
In the spirit of open science, all code and data analysis results can be found at Github.\footnote{\url{https://github.com/RichardHGL/CHI2025_Plan-then-Execute_LLMAgent}}

\subsection{Planning}
\label{sec-method-planning}
While LLMs can generate high-quality plans, there is no guarantee of their correctness and their further impact on the execution of the plan. 
Thus, involving users in the planning stage and controlling the plan quality would be essential to ensure successful subsequent execution. 

\begin{figure*}[htbp]
 \centering
  \subfigure[]
  {\label{fig:execution-flow-chart}
  \centering
  \includegraphics[height=0.42\textwidth]{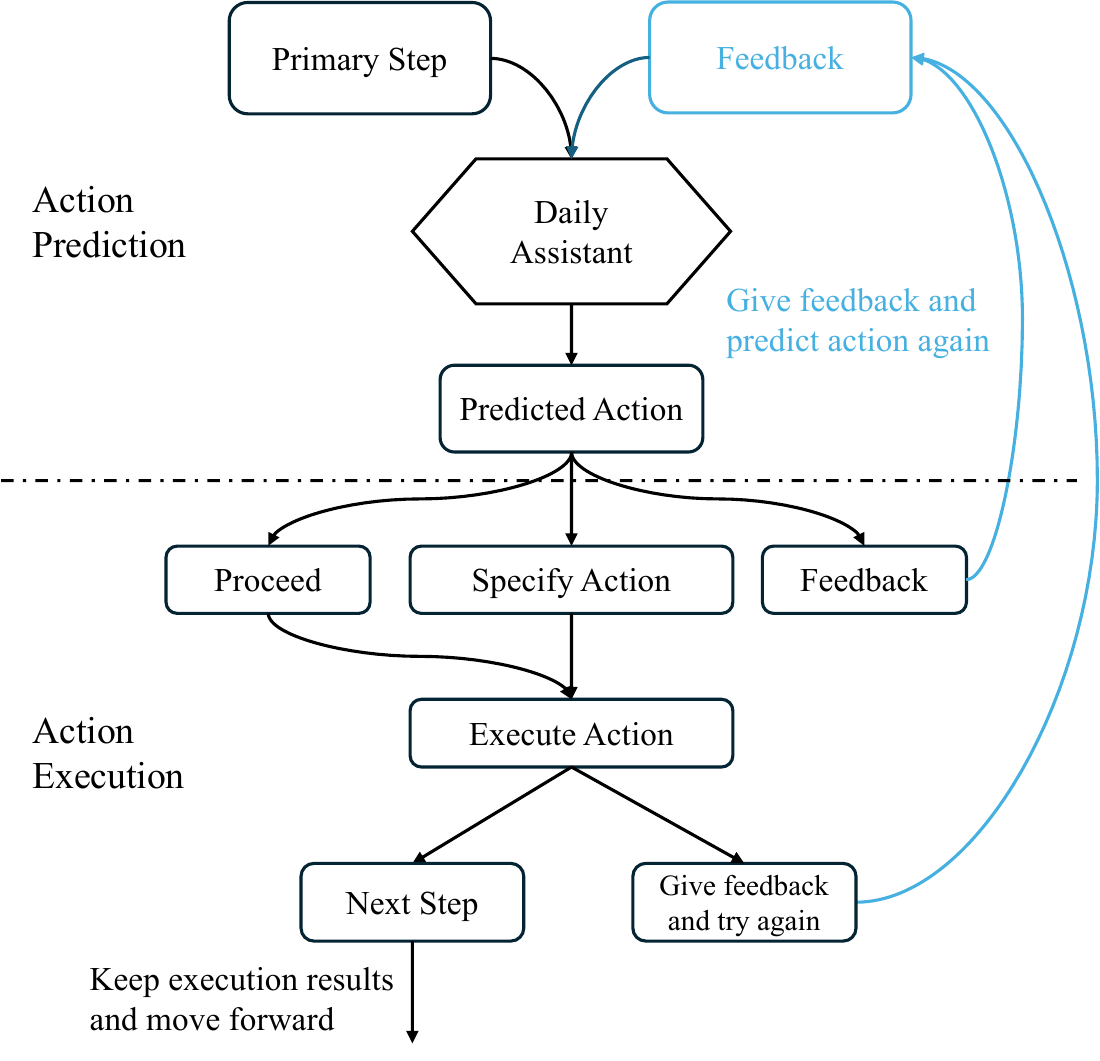}
 }
  \subfigure[]
  {\label{fig:execution-interface}
  \centering
  \includegraphics[height=0.42\textwidth,width=0.4\textwidth]{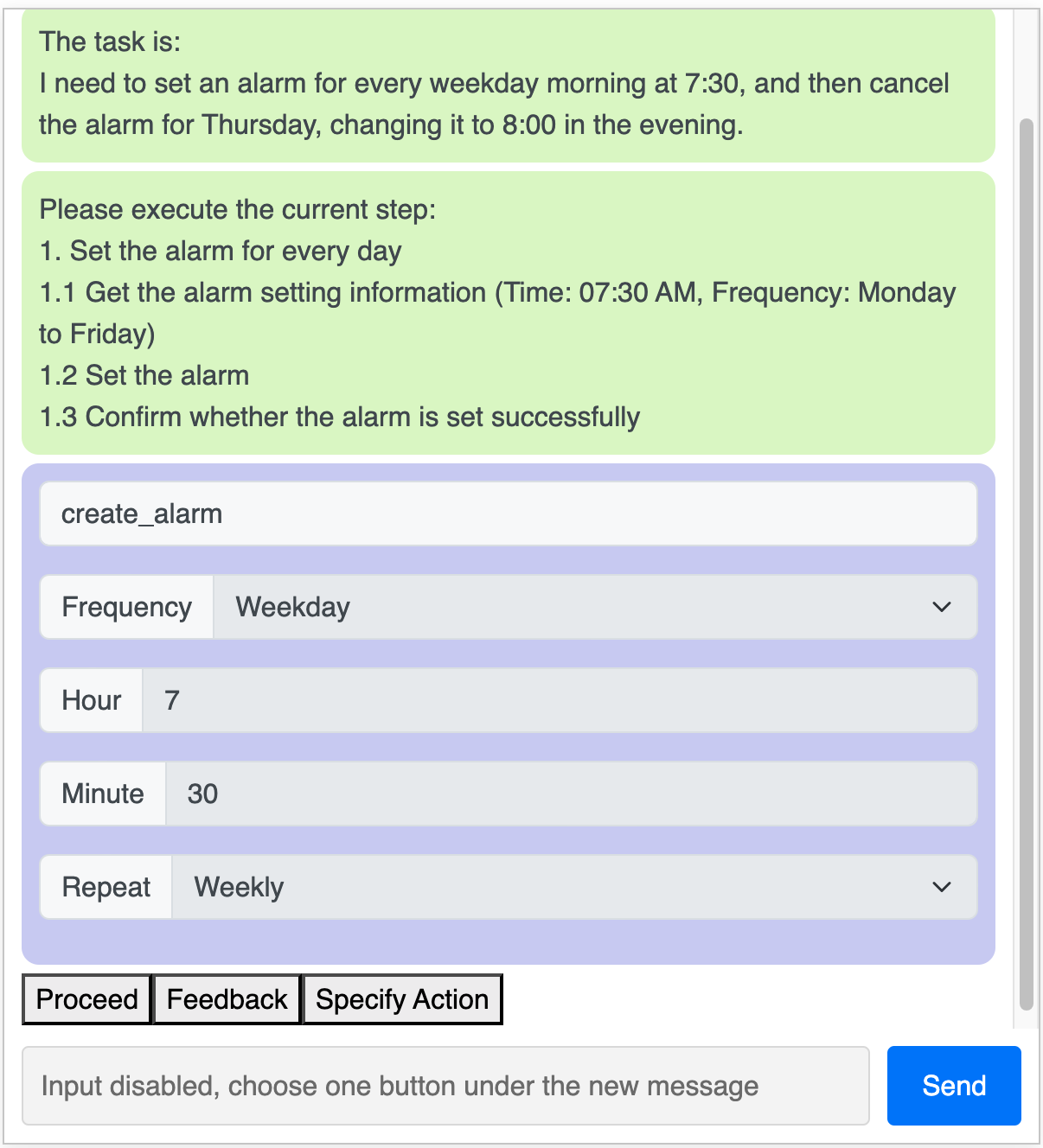}
 }
 \centering
 \caption{User-involved execution flow chart and interface. \revise{Panel (a): a flow chart illustrating how each primary step is executed with two stages: action prediction and action execution.} Panel (b): a screenshot of the conversation interface for user-involved execution.
 }
 \label{fig-interface}
 \Description{User-involved execution flow chart and interface. In the left panel, a flow chart shows how each primary step is executed with two stages: action prediction and action execution. In the right panel, we show one example of a user-involved execution page, which follows a conversation to show the step-by-step execution process. After one action is predicted, users can choose from options `Proceed', `Feedback', and `Specify Action' to approve or involve in the real-time execution.}
\end{figure*}

\paratitle{Plan Format}. 
The step-wise plan in our study followed a hierarchical structure, adapted from a benchmark for LLM agents toolkit usage~\cite{huang2024planning}. 
The whole plan consists of multiple sub-steps, which are at most three levels (\eg 1., 1.x, 1.x.y where x,y are integers). 
All sub-steps started with the same prefix index are denoted as one primary step (\eg the three blocks of planning outcome in Figure~\ref{fig:illustration}). 
A high-level step (\eg 1.) will provide high-level instruction of the current primary step, while low-level steps \revise{(\eg 1.x, 1.x.y)} will provide subsequent details. 
In the execution stage, each primary step will be used as the execution unit. 
The LLM agent will transform one primary step into a predicted action filled with parameters. 
Thus, we ask participants to provide all necessary details in sub-steps of each primary step. {Each primary step will be transformed into \textbf{single action} in the follow-up execution stage}. 
If one primary step requires two actions to accomplish, it may cause a potential loss of one action. 
Thus, when a plan contains one primary step that contains information about two potential actions (\eg the initial plan in Figure~\ref{fig:illustration}), we consider it as a low-quality plan with `grammar errors.'\footnote{Note that this is not to be confused with the notion of grammar in language.} 
All these plan format designs are informed in our onboarding tutorial.

\paratitle{User-involved Planning}. Figure~\ref{fig:planning} shows one screenshot of user-involved planning in our study. 
At the top of the interface, we provide a task description along with three buttons: `Show Potential Actions', `Plan Edit Instruction', and `Add one step'. 
By clicking `Show Potential Actions', we provide a prompt window to show concrete documentary descriptions of all potential actions (including action purpose and parameters) to be used in the execution stage. All instructions used in our tutorial are accessible with clicking the button `Plan Edit Instruction'. 
After users join the planning stage, an initial plan generated by LLM will be presented in the orange area. We allow users to edit the plan with following interactions:
\begin{itemize}
    \item Add step. By clicking `Add one step' button, users can insert a valid sub-step index into the whole plan, and then they can edit the plan text.
    \item Delete step. By clicking the `Delete step' button at the end of one step, all sub-steps associated with that step will be deleted from the plan.
    \item Edit step. By clicking the text input area in each step, users are allowed to edit the text with keyboard input.
    \item Split step. By clicking the `Split step' button associated with one step, we will split the original primary step into two primary steps. A new primary step will start the current step and contain all follow-up sub-steps. For example, if we click `Split step' for the plan show in Figure~\ref{fig:planning} at index `2.2'. We will generate a new blank step `3.' (where user input is expected) and re-index all sub-steps with `2.2.x' to `3.1.x'. At the same time, the original plan steps behind it will be automatically updated. Through this action, users can easily split one step that contains too much information into two primary steps. Figure~\ref{fig:illustration} shows an example of plan edit with `split step'.
\end{itemize}


\subsection{Execution}
\label{sec-method-execution}

After the planning stage, we obtain a plan with a step-wise structure. 
In the execution stage, the LLM agent executes the outcome of the planning stage \revise{(\ie a step-wise plan in text)} in a step-by-step manner. 
\revise{In each step, the LLM agent  translates a single step of the plan into one action, which is implemented with an API call in the backend. 
This setup is a simulation of real-world applications, which provide services with API calls (commonly implemented as langchain tools\footnote{\url{https://python.langchain.com/v0.1/docs/modules/tools/}}). Such a simulation setup is effective in developing and validating theory~\cite{davis2007developing} and has been widely adopted in existing research on agent-based modeling and HCI studies~\cite{olson2014ways}.} 
To provide a smooth user experience, we adopted a conversational interface to present the execution process. 
Figure~\ref{fig:execution-interface} shows one screenshot of user-involved execution in our study. 
As we can see, after a message of the first primary step of the plan, the LLM agent predicts one action `create\_alarm'. 
In our study, to provide a tidy view of the action prediction, we wrap the predicted action as one card (the blue area in Figure~\ref{fig:execution-interface}).

\paratitle{User-involved Execution}. 
Figure~\ref{fig:execution-flow-chart} \revise{presents} a flow-chart to illustrate a primary step executed by the daily assistant (\ie LLM agent). First, given one primary step, the daily assistant predicts an action based on a given list of prepared actions \revise{(\ie pre-defined APIs in the backend)}. 
After users check the predicted action, they can choose from \revise{one of the following} three buttons to respond. \textbf{(1)`Proceed':} 
It indicates users agree that the predicted action is correct. After clicking this button, the LLM agent moves forward to execute it and shows the execution result of this action. 
\textbf{(2) `Feedback':} Users can give text feedback based on the message input area at the bottom of the conversational interface. This  triggers another action prediction based on the current primary step and user feedback. Then, users are  provided with the three options \revise{to proceed} again. \textbf{(3) `Specify Action':} Users can override the current action prediction with the manual specification of one action. If users choose this response, they are first asked to choose one action from the prepared action list and then fill in the parameters manually. The LLM agent directly executes the user-specified action. 
After one action is executed, if users are not satisfied with the results, they can choose to re-execute this step by providing text feedback \revise{(\ie by clicking button `Give feedback and try again')}, which works similarly to the `Feedback' option. 
If users are satisfied with the execution results, they can click the `Next Step' button and move to execute the next primary step. 
By iterating over this process through the step-wise plan, users \revise{can choose to either} approve or get involved in modifying the execution outcomes in each step. \revise{All actions are predicted and executed in the backend (\ie the respective API calls are triggered).} 


\subsection{Hypotheses}
\label{sec:hypo}

Our experiment is designed to answer questions of how human involvement in the planning and execution stages will shape their trust and overall task performance. 
To analyze such impact, we regulate the levels of automation in the 
LLM agent through the planning and execution stage as baselines for comparison. 
The automatic planning and execution denotes that the LLM agent directly generates the task outcomes without user involvement.

With user involvement in the planning stage, users have the opportunity to fix potential mistakes or issues in the plan generated by LLMs. 
Working on such plan editing tasks is similar to debugging, which has been argued to bring about a critical mindset~\cite{he2024err} to the generated plan. 
With a critical mindset, users may better calibrate their trust in the planning outcome. 
We also consider user involvement in planning to be beneficial to the plan quality, which can then contribute to the overall task performance. 
Thus, we hypothesize that:
\begin{framed}
\noindent\textbf{(H1)}: Compared to automatic planning, user-involved planning will result in a higher calibrated  trust {in the plan}.\\
\noindent\textbf{(H2)}: Compared to automatic planning, user-involved planning will result in better overall task performance.
\end{framed}

In the user-involved execution process, users manually check the action prediction and execution results of each primary step. 
Such user involvement increases the chances of discovering potential mistakes of LLM agents.
Once users realize that the LLM agent made mistakes, they can get involved in modifying the execution outcome of the current step. 
By fixing these mistakes, the overall task performance gets improved. 
With such involvement in fixing potential errors, users will be more critical of trusting the task outcome. 
Therefore, we hypothesize that:
\begin{framed}
\noindent\textbf{(H3)}: Compared to automatic execution, user-involved execution will result in a {higher} calibrated trust {in execution outcome}.\\
\noindent\textbf{(H4)}: Compared to automatic execution, user-involved execution will result in better overall task performance.
\end{framed}

\section{Study Design}

This section describes our experimental conditions, tasks,  variables, procedure, and participants in our study. {Our
study was approved by the human research ethics committee of our institution.}

\subsection{Experimental Conditions}
In our study, users collaborate with LLM agent-based daily assistants in two stages: planning and execution. 
To comprehensively understand the effect of user involvement at each stage, we considered a 2 × 2 factorial design with four experimental conditions: (1) automatic planning, automatic execution (represented as AP-AE), (2) automatic planning, user-involved execution (represented as AP-UE), (3) user-involved planning, automatic execution (represented as UP-AE), (4) user-involved planning, user-involved execution (represented as UP-UE). 
In conditions with user-involved planning, users are allowed to edit the plan generated by LLM with the actions of edit/add/delete/split step. 
By comparison, in conditions with automatic planning, users will directly adopt the plan generated by the daily assistant.
In conditions with user-involved execution, users can interact with the step-by-step execution LLM agent (cf. Section~\ref{sec-method-execution}) and refine execution results with text feedback or manual specification. 
By comparison, in conditions with automatic execution, users will directly accept the automatic execution results.

\begin{table*}[htbp]
	\centering
	\caption{Selected tasks in our study. The `Risk' is based on the risk feedback obtained with pilot study. \#A and \#C refer to the number of actions and the number of named concepts in each task, respectively. }
	\label{tab:tasks}%
	\scalebox{.8}{
    \begin{tabular}{p{0.02\textwidth}|p{0.04\textwidth}|p{0.06\textwidth}|p{0.8\textwidth}|p{0.03\textwidth}|p{0.03\textwidth}| p{0.12\textwidth}}
		\hline
        \textbf{ID}&\textbf{Risk}& \textbf{Domain}&  \textbf{Task Description}& \textbf{\#A}& \textbf{\#C}& \textbf{Notes}\\
		\hline \hline
		1& High& Finance& My account ID is 54321, and the password is PWD2023. I plan to make two foreign exchange transactions. The first is to buy 10,000 euros (with USD), and the second is to sell 5,000 US dollars (to EUR). Please help me operate.&4&4& simple task, imperfect plan\\
        \hline
        2& High& Finance& Please inquire about the current debt amount of my credit card with the last five digits 12345, and deduct the corresponding 12000 USD from my savings card number 6212345678900011 to repay this debt, then help me check the amount of the outstanding bill for the same credit card within 30 days after today.&4&6& complex task, imperfect plan\\
        \hline
        3& High& Repair& I need to schedule a repair for my TV at 6 PM tomorrow evening. The brand is Sony, model X800H, and there is an issue with the screen. Please book the repair service and tell me the reservation number.&4&7& complex task, imperfect plan\\
        \hline
        4& Low& Alarm& I need to set an alarm for every weekday morning at 7:30, and then cancel the alarm for Thursday, changing it to 8:00 in the evening.&2&3& simple task, correct plan\\
        \hline
        5& Low& Flight& I have an important meeting to attend next Wednesday, and I need to book a flight ticket from London to Amsterdam for tomorrow, it must be a morning flight, and then return from Amsterdam to London tomorrow night, please handle it for me.&2&6& simple task, correct plan\\
        \hline
        6& Low& Travel& Please plan a trip for me departing on October 1st at 8:00 AM to Japan, returning on October 7th at 11:00 PM, including Tokyo Disneyland, Senso-ji Temple, Ginza, Mount Fuji, Kyoto cultural experience, Universal Studios Osaka, and visiting the Nara Deer Park on October 4th, and help me find hotels where the nightly cost does not exceed 10,000 Japanese yen.&3&11& complex task, correct plan\\
        \hline
            \hline
    \end{tabular}}
\end{table*}%

\subsection{Tasks}
To analyze how LLM agents can serve as daily assistants, we adopted tasks from a planning dataset designed for LLM agents --- UltraTool~\cite{huang2024planning}. 
We selected daily scenarios: currency transactions, credit card payments, repair service appointments, alarm setting, flight ticket booking, and trip itinerary planning. 
The selected tasks are shown in Table~\ref{tab:tasks}. \revise{For more details about how the plan-then-execute LLM agent works on the selected tasks (\eg automatic plan, pre-defined actions, automatic evaluation, and explanation for errors in automation), please refer to the appendix.}
All tasks in UltraTool dataset are annotated with the step-wise plan format described in Section~\ref{sec-method-planning}. \revise{The execution of these tasks is based on a simulation environment (described in Section~\ref{sec-method-execution}) where all required actions are implemented as backend APIs. 
In our study, all tasks are executed in a simulation setup, which has been a popular method for orchestrating meaningful human-centered AI studies~\cite{doshi2017towards,salimzadeh2024dealing}.} 

\paratitle{Task Selection}. First, based on the domain distribution of the UltraTool dataset, we selected seven domains: Finance, Alarm, Travel, Tracking, Restaurant, Flight, and Repair. 
For each domain, we only \revise{consider} tasks that contain more than ten steps (including all sub-steps) and require at least three uses of actions. 
Then, we manually selected ten tasks: four from the finance domain and one for each of the others. 
With a pilot study, we tested how users work on the ten tasks. 
We recruited 10 participants from the Prolific platform and only considered the feedback of 9 participants who passed all attention checks. 
\revise{Using} the question “How much risk do you perceive in this task when relying on this daily AI assistant?”, we collected the perceived risk of working with the LLM agents on each task \revise{using a 5-point Likert scale, ranging from \textit{1: not risky at all}---to---\textit{5:very risky}.} 
We categorize the ten tasks into a high-risk group (top 5) and a low-risk group (bottom 5). 
We selected three tasks from each group while balancing the complexity of the task description (three simple tasks and three complex tasks) and the correctness of the provided plan (three correct plans and three imperfect plans). 
\revise{Based on existing literature on task complexity~\cite{wood1986task,salimzadeh2023missing}, we  considered component complexity to inform our selection. 
This is assessed as the `total number of distinct information cues that need to be processed to perform the task'. 
Here, we considered the number of unique actions and the number of named concepts provided in each task.}
According to prior work~\cite{miller1956magical}, most people can only handle 5 to 9 concepts at the same time. 
The component complexity of all complex tasks in our study is more than nine.
The six tasks selected are shown in Table~\ref{tab:tasks}. 
Besides the six tasks, we used one simple task (\ie checking bank account balance) as the example in the onboarding tutorial.

\subsection{Measures and Variables}
\label{sec-measure}

The variables and measures used in our study refer to existing empirical studies of human-AI collaboration~\cite{lai2021towards}. 
All measures adopted in our study can be summarized in Table~\ref{tab:variables}. 
\begin{table*}[htbp]
	\centering
	\caption{The different variables considered in our experimental study. ``DV'' refers to the dependent variable.}
	\label{tab:variables}
	\begin{footnotesize}
	\begin{tabular}{c | c | c | c}
	    \hline
	    \textbf{Variable Type}&	\textbf{Variable Name}& \textbf{Value Type}& \textbf{Value Scale}\\
	    \hline \hline

	    \hline
         \multirow{2}{*}{Calibrated Trust (DV)}& Calibrated Trust in planning (CT$_p$)& Binary& 0: miscalibrated trust, 1: calibrated trust\\
        & Calibrated Trust in execution (CT$_e$)& Binary& 0: miscalibrated trust, 1: calibrated trust\\
        \hline
	    \multirow{3}{*}{Task Performance (DV)}& Plan Quality& Likert& 5-point, 1: low, 5: high\\
        & Action Sequence Accuracy (ACC$_s$)& Binary& 0: mismatch, 1: exact match with ground truth\\
        & Execution Accuracy (ACC$_e$)& Binary& 0: wrong execution result, 1: correct execution result\\
	    \hline
     \multirow{4}{*}{Trust}& Reliability/Competence& Likert& 5-point, 1: poor, 5: good\\
    & Understanding/Predictability& Likert& 5-point, 1: poor, 5: good\\
    & Intention of Developers& Likert& 5-point, 1: poor, 5: good\\
    & Trust in Automation& Likert& 5-point, 1:strong distrust, 5: strong trust\\
    \hline
     \multirow{4}{*}{Covariates}& LLM Expertise& Likert& 5-point, 1: No experience, 5: Extensive experience\\
     & Automatic Assistant Expertise& Likert& 5-point, 1: No experience, 5: Extensive experience\\
     & Propensity to Trust& Likert& 5-point, 1: tend to distrust, 5: tend to trust \\
     & Familiarity& Likert& 1: unfamiliar, 5: very familiar\\
     \hline
	\multirow{5}{*}{Exploratory}& Confidence& Likert& 5-point, 1: unconfident, 5: confident\\
    & Risk Perception& Likert& 5-point, 1: not risky at all, 5: very risky\\
    & Open Feedback on Planning& Text& Open Text\\
    & Open Feedback on Execution& Text& Open Text\\
    & Other Open Feedback& Text& Open Text\\
	    \hline
     \multirow{6}{*}{{Cognitive Load }}& Mental Demand& Likert& -7: very low, 7: very high\\
	    & Physical Demand& Likert& -7: very low, 7: very high\\
	    & Temporal Demand& Likert& -7: very low, 7: very high\\
	    & Performance& Likert& -7: Perfect, 7: Failure\\
        & Effort& Likert& -7: very low, 7: very high\\
        & Frustration& Likert& -7: very low, 7: very high\\
     \hline
	\end{tabular}
	\end{footnotesize}
\end{table*}

\paratitle{Calibrated Trust}. To assess calibrated trust in the planning stage and execution stage, we assessed user trust at each stage with a question ``Do you trust that [the execution of this plan / the execution process] can provide a correct outcome based on the task instructions?''. 
Based on the plan quality evaluation (5-point Likert), the calibrated trust in the planning (CT$_p$) is calculated based on the frequency at which users trusted the high-quality plan (expert annotation with 5) and users distrusted the plan with other evaluation results.
Similarly, for the calibrated trust in execution (CT$_e$), we calculated the frequency at which users trusted the correct execution results and distrusted the wrong execution results. 
The two measures can be calculated as:

\begin{equation}
\begin{aligned}
    \textnormal{CT}_p = &\mathbb{I}\left( \textnormal{trust = `Yes'}, \textnormal{plan quality} = 5\right) \\ & +\mathbb{I}\left( \textnormal{trust = `No'}, \textnormal{plan quality} < 5\right)
\end{aligned}
\end{equation}

\begin{equation}
    \textnormal{CT}_e = \mathbb{I}\left( \textnormal{trust = `Yes'}, \textnormal{ACC}_e = 1\right) + \mathbb{I}\left( \textnormal{trust = `No'}, \textnormal{ACC}_e = 0 \right)
\end{equation}

To assess the task performance, we mainly considered the task outcome from the planning and execution stages. 

\paratitle{Plan Quality}. As for the planning outcome, we evaluate the plan quality based on a 5-point Likert scale: 
1. low-quality plan, task requirements not covered; 
2. low-quality plan, task requirements covered but with grammar errors; 
3. medium-quality plan, task requirement covered but with at least one action intent mismatch with ground truth action sequence; 
4. medium-quality plan, task requirements covered but miss or have wrong details for action parameters; 
5. high-quality plan, covering all task requirements and providing all necessary details.

\paratitle{Execution Performance}. 
The execution of the step-wise plan will result in an action sequence. 
We provide a ground truth action sequence as a reference to evaluate the generated action sequence. 
We measure the action sequence accuracy (ACC$_s$) as the strict match of the action sequence and ground truth. 
Meanwhile, if one action sequence contains some redundant actions that are not harmful (\eg searching for flights), the execution results should still be correct. 
Thus, we also consider execution accuracy (ACC$_e$) as a task performance measure.

\paratitle{Subjective Trust and Covariates}. To enrich our analysis of user trust, we followed existing work to adopt the six subscales from the Trust-in-automation questionnaire~\cite{korber2019theoretical}. 
The four subscales --- \textit{Reliability/Competence},  \textit{Understanding/Predictability},  \textit{Intention of Developers},  \textit{Trust in Automation} are used as subjective measures of user trust in the LLM agent. 
Meanwhile, the \textit{Familiarity} and \textit{Propensity to Trust} are also used as covariates. Besides them, we considered user expertise in LLMs and user expertise in automatic assistants as covariates. 

\paratitle{Exploratory Variables}. To enrich our understanding of LLM agent as daily assistant, we assessed user confidence (both planning and execution) and risk perception along with each task. 
After users finish the study, we also ask for their open-text feedback on the planning and execution stages as well as other comments. 
To check the cognitive load of user involvement in our study, we adopted the NASA-TLX questionnaire~\cite{colligan2015cognitive}, which contains six subscales.


\subsection{Participants}

\paratitle{Sample Size Estimation}. 
To ensure sufficient statistical power, we estimated the required sample size for a 2 × 2 factorial design based on G*Power~\cite{faul2009statistical}. 
{To correct for testing multiple hypotheses, we applied a Bonferroni correction so that the significance threshold decreased to $\frac{0.05}{4}=0.0125$.} 
We specified the default effect size $f = 0.25$
(\textit{i.e.,} indicating a moderate effect), a significance threshold $\alpha = 0.0125$ (\textit{i.e.,} due to testing multiple hypotheses), a statistical power of $(1 - \beta) = 0.8$, and that we will investigate $4$ different experimental conditions/groups. 
This resulted in a required sample size of $244$ participants. 
We thereby recruited 347 participants from the crowdsourcing platform Prolific\footnote{\url{https://www.prolific.co}}, to accommodate potential exclusion.

\paratitle{Compensation}. All participants were rewarded with an hourly wage of \pounds 8.1 deemed to be ``\textit{Fair}'' payment by the platform (estimated completion time was 30 minutes). 
As participants in condition UP-UE spent longer in the study, we paid each participant a commensurate bonus accounting for an extra 10 minutes. 
We rewarded participants with extra bonuses of \pounds 0.05 for every high-quality plan and correct execution result. 
According to existing literature~\cite{lee2004trust}, such a bonus setup can help incentivize participants to reach a correct decision. 
\revise{In comparison with existing literature exploring human-AI decision making~\cite{lai2021towards}, our reward setup is above the average payment and can be considered as being sufficient to elicit ecologically valid behavior among participants (\ie aiming to arrive at accurate execution results). Moreover, similar bonus structures akin to our setup have been effective in incentivizing reliable participant behavior and improving data quality across different studies with crowdsourced participants~\cite{fan2020crowdco, salimzadeh2024dealing,liutilizing,ma2024you}.}

\begin{figure*}[h]
    \centering
    \includegraphics[width=\textwidth]{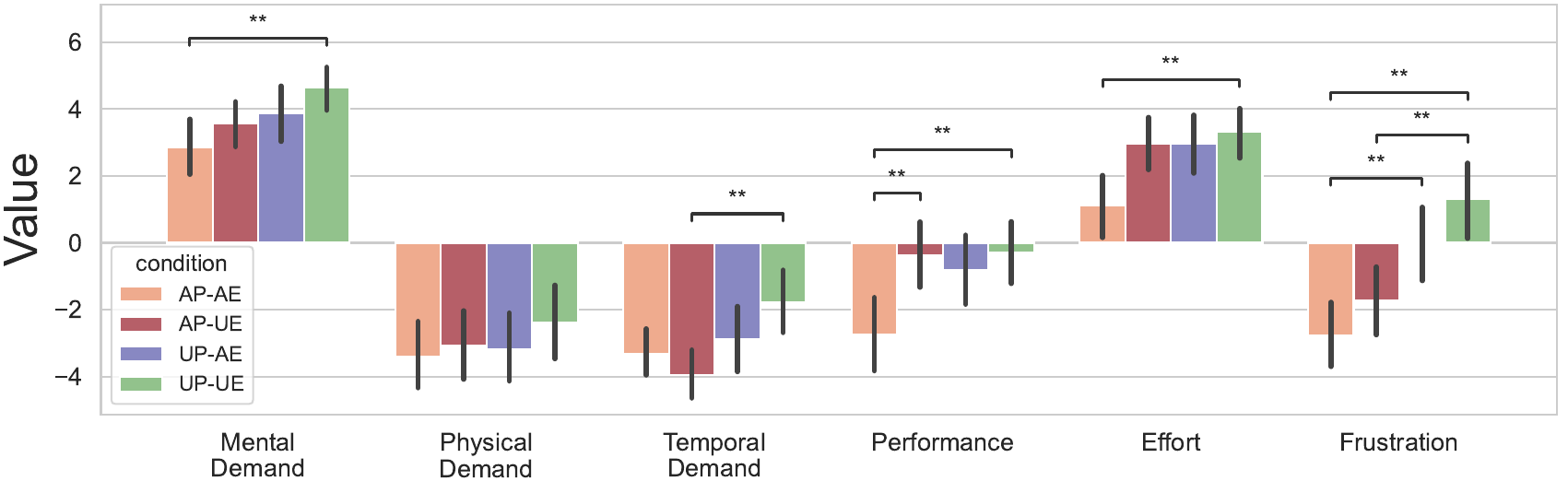}
    \caption{\revise{Bar plot for cognitive load across all conditions. ** indicates significance ($p < 0.0125$) through post-hoc Tukey HSD test. The error bars represent the 95\% confidence interval.}}
    \label{fig:cognitive_load}
    \Description{Bar plot for cognitive load across all conditions. User involvement in planning shows a significant impact on Mental Demand, Temporal Demand, and Frustration. User involvement in execution shows a significant impact on Performance and Effort. These results indicate that user involvement in planning and execution will require a relatively high cognitive load.}
\end{figure*}

\paratitle{Filter Criteria}. All participants were proficient English speakers between the ages of 18 - 50. 
We also constrained their prior experience (at least 40 successful submissions) and had an approval rate of above 90\% on the Prolific platform. 
We excluded participants from our analysis if they failed any attention check, or represented an outlier regarding the plan quality. 
Outliers were 4 participants who generated more than three low-quality plans among six tasks. 
The reserved 248 participants had an average age of 32.5 ($SD$= 8.1) and a balanced gender distribution
($50\%$, $49.6\%$ female, $0.4\%$ other).


\subsection{Procedure}
At the beginning of our study, we showed informed consent for data collection and the study's purpose. 
Only participants who signed the informed consent \revise{were allowed to} continue to work on our study. 
Next, participants were asked to complete a pre-task questionnaire to measure their expertise in LLM and automatic assistants. 

Participants were then assigned to one of the experimental conditions, which differed in the level of user involvement in the planning stage and execution stage. 
With an onboarding tutorial, we showcased the necessary interactions that participants were expected to perform in the planning and execution stages. 
We used an example task to help participants understand how to work with the plan-then-execute LLM agent. 
After the onboarding tutorial, participants worked on the selected tasks, which were shuffled at random for every participant to prevent task ordering effects. 
After the participants finished the task batch, we measured their perceived cognitive load using the NASA-TLX questionnaire~\cite{colligan2015cognitive}, their overall trust in the daily assistant using the trust in automation questionnaire~\cite{korber2019theoretical}, and we gathered their feedback on our system (related to planning, execution, and other aspects) using open-ended text.
\section{Results} 
In this section, we will present the main experimental results and exploratory analysis for our study. 

\subsection{Descriptive Statistics}

In total, our analysis is based on 248 participants, who are balanced across conditions: AP-AE (63), AP-UE (64), UP-AE (61), and UP-UE (60). 
All edited plans in user-involved planning conditions are evaluated by the authors following the plan quality criteria \revise{described} in Section~\ref{sec-measure}. 

\paratitle{Distribution of Covariates}. In our study, most participants claimed to have some experience with using large language models ($M=3.6, SD=1.0$) and automatic assistants ($M=3.4, SD=1.1$). In the trust in automation questionnaire, participants indicated a medium level of \textit{Familiarity} ($M=2.9, SD=1.2$) and \textit{Propensity to Trust} ($M=3.0, SD=0.7$).

\paratitle{Performance Overview}. Overall, users show calibrated trust in the planning ($M= 0.50, SD=0.13$) and calibrated trust in the execution ($M=0.64, SD=0.19$). 
For the execution outcome, we find that although it is tricky to obtain a ground truth action sequence ($M=0.48, SD=0.17$), the action sequence has a relatively high recall of ground truth actions ($M=0.77, SD=0.11$). 
The successful rate for correct execution ($M=0.52, SD=0.18$) is higher than the strict evaluation of the action sequence. 
We also collected user subjective trust with four subscales of the trust in automation questionnaire: \textit{Reliability/Competence} ($M=3.49, SD=0.77$),  \textit{Understanding/Predictability} ($M=3.30, SD=0.56$),  \textit{Intention of Developers} ($M=3.61, SD=0.81$),  \textit{Trust in Automation} ($M=3.52, SD=1.01$). 
With a two-way ANOVA analysis considering user involvement in planning and execution, we do not find any significant impact of user involvement on subjective user trust in AI systems across conditions.


\paratitle{Cognitive Load}. The cognitive load of participants across the four experimental conditions is shown in Figure~\ref{fig:cognitive_load}. 
Based on two-way ANOVA, we analyzed the impact of user involvement in planning and execution affect user cognitive load. 
User involvement in planning shows a significant impact on \textit{Mental Demand}, \textit{Temporal Demand}, and \textit{Frustration}. User involvement in execution shows a significant impact on \textit{Performance} and \textit{Effort}. 
With post-hoc Tukey HSD test, we confirmed such impact --- involvement in both planning and execution  posed a higher cognitive load on participants.


\begin{table*}[h]
	\centering
	\caption{Task-specific evaluation results for user-involvement in planning on calibrated trust in planning (CT$_p$) and plan quality. We also report the mean value for each measure on each condition.}
	\label{tab:h1-res}
    \begin{small}
	\begin{tabular}{c | c c c  c | c | c  c c c | c}
	    \hline
        &     \multicolumn{5}{c|}{CT$_p$}&  \multicolumn{5}{c}{Plan Quality}\\
        \cline{2-11}
        Tasks& AP-AE& AP-UE& UP-AE& UP-UE& Post-hoc results& AP-AE& AP-UE& UP-AE& UP-UE& Post-hoc results\\
        \hline 
        \hline
        Avg & 0.51& 0.50& 0.50& 0.50& - & 3.8& 3.8& 3.6& 3.7& AP > UP\\
        \hline
        task-1& 0.11 & 0.20 & 0.13 & 0.27& -& 2.0 & 2.0 & 2.3 & 2.4& AP < UP\\
        task-2 &0.21 & 0.11 & 0.20 & 0.17& -& 3.0 & 3.0 & 2.9 & 2.9& -\\
        task-3 & 0.10 & 0.03 & 0.10 & 0.07& -& 3.0 & 3.0 & 2.7 & 2.9& AP > UP\\
        task-4 & 0.94 & 0.97 & 0.80 & 0.90& AP > UP& 5.0 & 5.0 & 4.3 & 4.8& AP > UP\\
        task-5 & 0.87 & 0.84 & 0.90 & 0.82& -& 5.0 & 5.0 & 4.6 & 4.8& AP > UP\\
        task-6 & 0.81 & 0.81 & 0.85 & 0.75& -& 5.0 & 5.0 & 4.7 & 4.6& AP > UP\\
    \hline
	\end{tabular}
 \end{small}
\end{table*}

\begin{table*}[h]
	\centering
	\caption{Task-specific evaluation results for user-involvement in planning on task performance. ACC$_s$ denotes the strict accuracy of an action sequence, and ACC$_e$ denotes the correctness of execution results. Bold fonts are used to highlight the best performance across conditions.}
	\label{tab:h2-res}
    \begin{small}
	\begin{tabular}{c | c c c  c | c | c  c c c | c}
	    \hline
        \multirow{2}{*}{Tasks}&     \multicolumn{5}{c|}{ACC$_s$}&  \multicolumn{5}{c}{ACC$_e$}\\
        \cline{2-11}
        & AP-AE& AP-UE& UP-AE& UP-UE& Post-hoc results& AP-AE& AP-UE& UP-AE& UP-UE& Post-hoc results\\
        \hline 
        \hline
        Avg& \textbf{0.53} & 0.46 & 0.46 & 0.48 & - & 0.54 & 0.53 & 0.47 & \textbf{0.56} & -\\
        \hline
        task-1& 0.00 & 0.00 & 0.10 & \textbf{0.12} & AP < UP & 0.00 & 0.00 & 0.10 & \textbf{0.13} & AP < UP\\
        task-2 &\textbf{0.78} & 0.64 & 0.61 & 0.57 & - & \textbf{0.78} & 0.72 & 0.66 & 0.75 & - \\
        task-3 & \textbf{0.44} & 0.12 & 0.36 & 0.28 & - & 0.44 & 0.42 & 0.36 & \textbf{0.52} & - \\
        task-4 & \textbf{0.95} & 0.89 & 0.75 & 0.82 & AP > UP & \textbf{0.95} & 0.89 & 0.75 & 0.82 & AP > UP \\
        task-5 & \textbf{0.98} & 0.91 & 0.90 & 0.90 & - & \textbf{0.98} & 0.91 & 0.92 & 0.90 & - \\
        task-6 & 0.05 & \textbf{0.22} & 0.02 & 0.18 & - & 0.06 & \textbf{0.23} & 0.03 & 0.22 & - \\
    \hline
	\end{tabular}
 \end{small}
\end{table*}

\paratitle{User Involvement}. Among 121 participants in conditions with user-involved planning, 104 participants edited at least one task plan. Meanwhile, 90 participants used the provided buttons (\ie add/delete/split step) in our study. In total, \textit{delete step} is used 394 times, \textit{add step} is used 183 times, \textit{split step} is used 126 times. Among 124 participants in conditions with user-involved execution, 114 participants interacted with the conversation interface to change action prediction (\ie have at least one task where they choose to give feedback or override predicted action). 
Meanwhile, 105 participants specified at least one action in the task batch. 
In total, \textit{Specify Action} is used 445 times, feedback to the LLM agent is used 91 times before action execution, and feedback to the LLM agent is used 163 times after execution.

\subsection{Hypothesis Verification}
As the tasks selected in our study are of different initial plan quality and risk levels, we conducted a task-specific analysis in each hypothesis verification.

\subsubsection{The Impact of User Involvement in Planning on Calibrated Trust} 
To verify \textbf{H1}, we adopted the one-way ANOVA test and post-hoc Tukey HSD test on the calibrated user trust in planning (\ie CT$_p$). 
The results are shown in Table~\ref{tab:h1-res}. 
Only in task-4, we found user involvement in planning will have a negative impact on calibrated trust in planning. 
To avoid a potential impact of user involvement in the execution stage, we conducted a two-way ANOVA test to confirm the findings. 
We only find a significant difference in task-4.
Post-hoc Tukey HSD results show that participants in conditions with automatic planning (AP) showed significantly higher calibrated trust in planning outcomes than those in conditions with user-involved planning (UP). 
Thus, our experimental results do not support \textbf{H1}.


We noticed that the calibrated trust in planning is quite low in the high-risk tasks where all initial plans are imperfect. 
This indicates that many users across all conditions consider the generated plan trustworthy. 
On tasks with low risk, where the initial plan is of high quality, users achieved much higher calibrated trust in the planning outcome. 
We also find that conditions with user-involved execution (UE) show slightly higher CT$_p$ in task-1 and task-4 than conditions with automatic execution (AE). With the same statistical test as \textbf{H1} analysis, such differences are not significant.

\subsubsection{The Impact of User Involvement in Planning on Task Performance} \label{sec-exp-h2}
To verify \textbf{H2}, we considered plan quality, the accuracy of action sequences (ACC$_s$), and the execution accuracy of the plan (ACC$_e$) for analysis.
For plan quality (cf. Table~\ref{tab:h1-res}), we conducted one-way ANOVA on plan quality considering the user involvement in the planning stage. 
We found that overall user involvement in the planning stage caused a decrease plan quality, especially on tasks with a perfect plan (\ie task 4, 5, 6, where plan quality = 5) and task-3. 
However, in task-1, where the original plan contains a grammar error, we find that user involvement in planning can improve the plan quality. 
As the action sequence accuracy (ACC$_s$) and execution accuracy (ACC$_e$) are not normally distributed, we conducted the Kruskal-Wallis H-test by considering the user involvement in the planning as the independent variable. \revise{The results are shown in Table~\ref{tab:h2-res}.}
With further post-hoc Mann-Whitney tests, we found that while participants achieved a relatively higher accuracy of action sequences in condition AP-AE, the condition UP-UE achieved the best execution accuracy. 
In most tasks, condition UP-UE achieved better or compatible performance as other conditions. The only exception is task-4, where user involvement in the planning caused a significantly worse performance (both ACC$_s$ and ACC$_e$). 
\revise{As user involvement does not consistently lead to improved performance, these results are not enough to support \textbf{H2}.}

\revise{We found that in task-1 and task-6 most participants in the AP-AE condition achieved a very low success rate. This is mainly due to the imperfect plans and imperfect execution generated by LLMs. 
In task-1, the plan generated by LLMs includes one step which contains two actions to execute. 
Due to the inability to edit the plan, the LLM agent execution missed one transaction in conditions with automatic planning. 
In task-6, the plan generated by LLMs is correct. 
However, in the automatic execution of step 2 of the plan (\ie selecting an itinerary suggested), the LLM agent has a high probability of choosing an itinerary that does not match the task description. 
If the participants do not carefully check the task description, and correct this agent behavior, the execution results would be wrong.
This also helps explain why user involvement substantially improves the task outcome accuracy in task-6. More details about tasks can be found in the appendix.}

\begin{table*}[h]
	\centering
	\caption{Task-specific evaluation results for user-involvement in execution on task performance. Bold fonts are used to highlight the best performance across conditions.}
	\label{tab:h4-res}
    \begin{small}
	\begin{tabular}{c | c c c  c | c | c  c c c | c}
	    \hline
        \multirow{2}{*}{Tasks}&     \multicolumn{5}{c|}{ACC$_s$}&  \multicolumn{5}{c}{ACC$_e$}\\
        \cline{2-11}
        & AP-AE& AP-UE& UP-AE& UP-UE& Post-hoc results& AP-AE& AP-UE& UP-AE& UP-UE& Post-hoc results\\
        \hline 
        \hline
        Avg& \textbf{0.53} & 0.46 & 0.50 & 0.51 & - & 0.54 & 0.53 & 0.50 & \textbf{0.58} & -\\
        \hline
        task-1& 0.00 & 0.00 & 0.10 & \textbf{0.12} & - & 0.00 & 0.00 & 0.10 & \textbf{0.14} & -\\
        task-2 &\textbf{0.78} & 0.64 & 0.67 & 0.62 & - & \textbf{0.78} & 0.72 & 0.69 & \textbf{0.78} & - \\
        task-3 & \textbf{0.44} & 0.12 & 0.42 & 0.29 & AE > UE & 0.44 & 0.42 & 0.42 & \textbf{0.53} & - \\
        task-4 & \textbf{0.95} & 0.89 & 0.94 & 0.88 & - & \textbf{0.95} & 0.89 & 0.94 & 0.88 & - \\
        task-5 & 0.98 & 0.91 & \textbf{1.00} & 0.98 & - & 0.98 & 0.91 & \textbf{1.00} & 0.98 & -\\
        task-6 & 0.05 & \textbf{0.22} & 0.02 & 0.19 & AE < UE & 0.06 & \textbf{0.23} & 0.04 & \textbf{0.23} & AE < UE\\
    \hline
	\end{tabular}
 \end{small}
\end{table*}

\subsubsection{The Impact of User Involvement in Execution on Calibrated Trust in Execution Outcome}
As we observe in Table~\ref{tab:h1-res}, user involvement in planning can have some negative impact on the plan quality, which further impacts the execution stage. 
To control such impact, we filtered out the tasks where plan quality decreased after user-involved planning in the analysis of user involvement in the execution stage. 
To verify \textbf{H3}, we conducted one-way ANOVA on calibrated trust in execution outcome (CT$_e$). The results are shown in Table~\ref{tab:h3-res}. 
We found that user involvement in execution causes no significant difference across conditions. 
Thus, \textbf{H3} is not supported by our experimental results.

\begin{table}[h]
	\centering
	\caption{Task-specific evaluation results for user-involvement in execution on calibrated trust in execution (CT$_e$). We also report the mean value for each measure on each condition.}
	\label{tab:h3-res}
    \begin{small}
	\begin{tabular}{c | c c c  c | c }
	    \hline
        \multirow{2}{*}{Tasks}&     \multicolumn{5}{c}{CT$_e$}\\
        \cline{2-6}
        & AP-AE& AP-UE& UP-AE& UP-UE& Post-hoc results\\
        \hline 
        \hline
        Avg & \textbf{0.66} & 0.65 & 0.64 & 0.65 & - \\
        \hline
        task-1& 0.48 & 0.44 & \textbf{0.51} & 0.49 & -\\
        task-2 &0.78 & \textbf{0.83} & 0.71 & 0.80 & -\\
        task-3 & 0.51 & 0.41 & \textbf{0.60} & 0.47 & -\\
        task-4 & \textbf{0.94} & 0.92 & 0.88 & 0.86 & -\\
        task-5 & 0.89 & 0.92 & \textbf{0.96} & 0.94 & -\\
        task-6 & 0.37 & 0.38 & 0.28 & \textbf{0.42} & -\\
    \hline
	\end{tabular}
 \end{small}
\end{table}


\subsubsection{The Impact of User Involvement on Overall Task Performance} Similar to the verification of \textbf{H3}, we excluded the tasks where plan quality decreased after user-involved planning in this analysis. 
As the plan is generated before user involvement in the execution, we only considered ACC$_s$ and ACC$_e$ in the analysis of user involvement in the execution stage. 
To verify \textbf{H4}, we conducted Kruskal-Wallis H-test by considering the user involvement in the execution as the independent variable. The results are shown in Table~\ref{tab:h4-res}. 
With post-hoc Mann-Whitney tests, we found that user involvement in the execution stage showed significantly higher ACC$_s$ and ACC$_e$ in task-6 (where the LLM assistant mainly failed to choose the most suitable itinerary plan). 
\revise{We found that participants in the AP-AE condition achieved the best accuracy of action sequences (\ie ACC$_s$), and participants in condition UP-UE achieved the best execution accuracy (\ie ACC$_e$). 
In other words, the executed action sequence in condition AP-AE is more aligned with the ground truth action sequence annotated by the authors. 
However, with user involvement in the execution stage, participants in condition UP-UE have a better opportunity to obtain correct task outcomes by correcting potentially flawed actions. 
Such a difference is due to our measure of ACC$_e$, which tolerates the non-risky actions (\eg search flight) and failure of action predictions. 
In contrast, our measure of ACC$_s$ considers this as a wrong action sequence. 
} 
\revise{Thus,} in task-3, \revise{even if} we find automatic execution achieved \revise{significantly} better ACC$_s$ than user-involved execution, participants in condition AP-UE and UP-UE 
obtained comparable or higher execution accuracy \revise{(\ie ACC$_e$) than conditions with automatic execution}. 
\revise{While user involvement shows some positive impact on the execution accuracy, such impact is not significant and consistent across all tasks. 
Only in task-6, where users can correct the errors made by the LLM agent (\ie the wrong itinerary selection mentioned in Section~\ref{sec-exp-h2}), user involvement in the execution shows a significant contribution to the task performance. 
Thus, these results are not enough to strictly support \textbf{H4}.
}

\subsection{Exploratory Analysis}

\subsubsection{The Impact of Covariates}
For further insights into all user factors on user trust and team performance, we calculated Spearman rank-order correlation coefficients for user trust, calibrated trust, risk perception, and task performance. 
As can be seen in Table~\ref{tab:correlation}, we found these covariates mainly show correlations with subjective user trust, calibrated trust in execution, and risk perception. 
First, all covariates (\ie user factors) positively correlated with user trust (four subscales in the trust in automation questionnaire~\cite{korber2019theoretical}) and negatively correlated with perceived risk (average over six tasks). 
It indicates that users with more expertise or familiarity with such systems tend to trust the daily assistant and show less perceived risk when using it. 
Meanwhile, users with a general propensity to trust also tend to trust the AI system. 
Besides user trust, \textit{Assistant Expertise} and \textit{Propensity to Trust} show a significant negative correlation with calibrated trust in the execution outcome. 
Apart from the above correlation, these user factors do not significantly correlate with task performance measures or calibrated trust in the planning outcome.

\begin{table*}[ht]
	\centering
	\caption{Spearman rank-order correlation coefficient for covariates level on dependent variables. All measures are calculated based on average over task batch. ``${\dagger}$'' and ``${\dagger\dagger}$'' indicate the effect of the variable is significant at the level of 0.05 and 0.0125, respectively.}
	\label{tab:correlation}%
	\begin{small}
        \scalebox{0.95}{
	\begin{tabular}{l | l| c c | c c | c c | c c}
	    \hline
	    \multicolumn{2}{c|}{\textbf{Covariates}}& \multicolumn{2}{c|}{llm expertise} &  \multicolumn{2}{c|}{assistant expertise}& \multicolumn{2}{c|}{Familiarity}& \multicolumn{2}{c}{Propensity to Trust}\\
     \hline
     \textbf{Category} &\textbf{Variables}&  $r$& $p$ &  $r$& $p$ &  $r$& $p$ &  $r$& $p$ \\
	    \hline \hline
    \multirow{4}{*}{\textbf{User Trust}}& Reliability/Competence & 0.334 &\textbf{.000}$^{\dagger\dagger}$& 0.245 &\textbf{.000}$^{\dagger\dagger}$& 0.321 &\textbf{.000}$^{\dagger\dagger}$& 0.679 &\textbf{.000}$^{\dagger\dagger}$\\
    &Understanding/Predictability & 0.307 &\textbf{.000}$^{\dagger\dagger}$& 0.164 &\textbf{.010}$^{\dagger\dagger}$& 0.208 &\textbf{.001}$^{\dagger\dagger}$& 0.380 &\textbf{.000}$^{\dagger\dagger}$\\
    &Intention of Developers & 0.406 &\textbf{.000}$^{\dagger\dagger}$& 0.324 &\textbf{.000}$^{\dagger\dagger}$& 0.362 &\textbf{.000}$^{\dagger\dagger}$& 0.517 &\textbf{.000}$^{\dagger\dagger}$\\
    &Trust in Automation & 0.380 &\textbf{.000}$^{\dagger\dagger}$& 0.278 &\textbf{.000}$^{\dagger\dagger}$& 0.356 &\textbf{.000}$^{\dagger\dagger}$& 0.698 &\textbf{.000}$^{\dagger\dagger}$\\
	    \hline
    \multirow{2}{*}{\textbf{Calibrated Trust}}&CT$_p$ & 0.053 & .404& 0.053 & .402& 0.056 & .378& 0.037 & .566\\
    &CT$_e$ & -0.120 & .059& -0.195 &\textbf{.002}$^{\dagger\dagger}$& -0.032 & .621& -0.174 &\textbf{.006}$^{\dagger\dagger}$\\
    \hline
    \textbf{Risk Perception}& Perceived Risk & -0.187 &\textbf{.003}$^{\dagger\dagger}$& -0.180 &\textbf{.004}$^{\dagger\dagger}$& -0.237 &\textbf{.000}$^{\dagger\dagger}$& -0.363 &\textbf{.000}$^{\dagger\dagger}$\\
    \hline
    
    \multirow{3}{*}{\textbf{Task Performance}}&ACC$_s$ & 0.037 & .560& -0.014 & .823& 0.110 & .085& 0.018 & .772\\
    &ACC$_e$ & -0.000 & .995& -0.037 & .567& 0.085 & .184& 0.007 & .911\\
    &Plan Quality& -0.035 & .587& -0.037 & .560& 0.080 & .211& -0.032 & .611\\
    \hline
	\end{tabular}
        }
	\end{small}
\end{table*}

\begin{table}[ht]
	\centering
	\caption{Task-specific spearman rank-order correlation coefficient for plan quality and risk perception. ``${\dagger}$'' and ``${\dagger\dagger}$'' indicate the effect of the variable is significant at the level of 0.05 and 0.0125, respectively.}
	\label{tab:correlation-2}%
	\begin{small}
        \scalebox{0.95}{
	\begin{tabular}{l | l| c c | c c }
	    \hline
	    \multirow{2}{*}{\textbf{Category}}&\multirow{2}{*}{\textbf{Variables}}& \multicolumn{2}{c|}{Plan Quality} &  \multicolumn{2}{c}{Risk Perception}\\
     \cline{3-6}
     & &  $r$& $p$ &  $r$& $p$\\
	    \hline \hline
    \multirow{2}{*}{\textbf{User Trust}}&Trust-p & 0.056 & .032$^{\dagger}$& -0.293 &\textbf{.000}$^{\dagger\dagger}$\\
    &Trust-e & 0.258 &\textbf{.000}$^{\dagger\dagger}$& -0.160 &\textbf{.000}$^{\dagger\dagger}$\\
    \hline
    \multirow{2}{*}{\textbf{Calibrated Trust}}& CT$_p$ & 0.723 &\textbf{.000}$^{\dagger\dagger}$& -0.102 &\textbf{.000}$^{\dagger\dagger}$\\
    & CT$_e$& 0.221 &\textbf{.000}$^{\dagger\dagger}$& 0.000 & .995\\
    \hline
    \multirow{3}{*}{\textbf{Task Performance}}&Plan Quality & - &-& -0.141 &\textbf{.000}$^{\dagger\dagger}$\\
    &ACC$_e$ & 0.400 &\textbf{.000}$^{\dagger\dagger}$& -0.110 &\textbf{.000}$^{\dagger\dagger}$\\
    &ACC$_s$ & 0.446 &\textbf{.000}$^{\dagger\dagger}$& -0.096 &\textbf{.000}$^{\dagger\dagger}$\\
    \hline
    \multirow{2}{*}{\textbf{Confidence}}&Confidence-p & 0.137 &\textbf{.000}$^{\dagger\dagger}$& -0.532 &\textbf{.000}$^{\dagger\dagger}$\\
    &Confidence-e & 0.225 &\textbf{.000}$^{\dagger\dagger}$& -0.271 &\textbf{.000}$^{\dagger\dagger}$\\
    \hline
	\end{tabular}
        }
	\end{small}
\end{table}

\subsubsection{Impact of Plan Quality and Risk Percetion.} Besides the measures calculated over task batch, a task-level 
analysis of plan quality and risk perception can deepen our understanding of their impacts. 
Besides measures adopted in Table~\ref{tab:correlation}, we include task-level confidence in this analysis and exclude the subscales from the trust in automation questionnaire.
Thus, we calculated Spearman rank-order correlation coefficients for task-level measures across all groups of participants (shown in Table~\ref{tab:correlation-2}). 
As we can see, both plan quality and risk perception significantly correlate with user trust, calibrated trust, task performance, and user confidence. 
The \textit{plan quality} shows a significant positive correlation with most measures, which indicates users perform better and calibrate their trust in the LLM agents in tasks with a high-quality plan. 
By contrast, the \textit{risk perceptions} shows a negative correlation with most measures and also a negative correlation with the plan quality. 

\subsubsection{Failure Analysis}
\label{sec-failure-analysis}
As we find that plan quality substantially affects task execution accuracy, we look into task performance across different plan qualities. 
For the tasks with low-quality plans (plans fail to cover task information or plan with grammar errors, \ie plan quality=1, 2), the execution accuracy is $1.8\%$. 
While for tasks with a plan that may mislead action prediction (plan quality = 3, 4), our LLM agent-based daily assistant achieved $59\%$ execution accuracy. 
The average execution accuracy for tasks with a high-quality plan (plan quality =5) is $66.7\%$.

We further check 717 tasks where a high-quality plan (plan quality = 5) is provided. Among them, 235 tasks provide wrong execution results. The main causes are: 
(1) Wrong action parameter prediction ($48.9\%$). 
While action names match, one or more parameters mismatch the expected value at some step of the action sequence. 
(2) Invalid actions ($48.5\%$). Given a perfect plan, the LLM agent failed to predict one valid action (failed to predict one action name or failed to predict some action parameter value) to execute in some steps. 
(3) Wrong action name prediction ($2.6\%$). The generated action sequence has at least one action name prediction that mismatches the ground truth. 

\begin{figure}[h]
    \centering
    \includegraphics[width=0.48\textwidth]{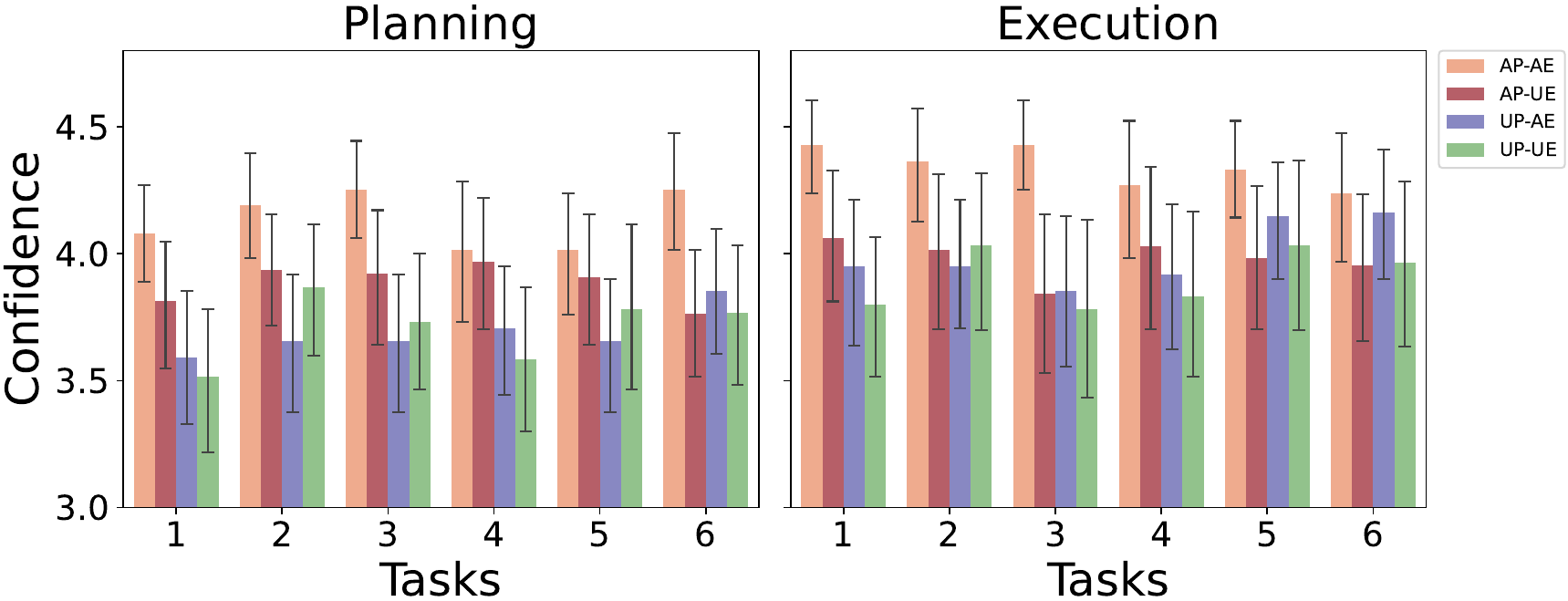}
    \caption{\revise{Bar plot for confidence dynamics, the x-axis denotes the task ordering index (shuffled for every participant). The error bars represent the 95\% confidence interval.}} 
    \label{fig:confidence}
    \Description{Bar plot for confidence dynamics, the x-axis denotes the task ordering index (shuffled for every participant). We observe that participants show higher confidence in the task execution outcomes, compared to planning outcomes. The confidence gap between the two stages get narrowed with user involvement in the execution stages.}
\end{figure}

\begin{table*}[htbp]
	\centering
	\caption{Excerpts from participants' responses to open questions regarding opinion.}
	\label{tab:example-opinion}%
	\scalebox{.9}{
    \begin{tabular}{p{0.72\textwidth}| p{0.08\textwidth} |p{0.24\textwidth}}
		\hline
		\textbf{Opinion towards Planning} &  \textbf{Sentiment}& \textbf{Reason}  \\
		\hline \hline
		I really like how organized it is. The step by step and numerical planning allows it to make sense in a clear and structured way, meaning there is less room for errors or misinformation& Positive&  helpful with reducing error\\
        \hline
        It was remarkable how quickly. It was able to achieve the goals which was set out in the tasks. I quite liked it I would definitely want something like this in my life as It would my my life much easier& Positive& Effective and make life easy\\
		\hline
		As I said previously, it's far, far too detailed in an unnecessary way.  I'm not sure people need the entire plan of what the AI will do, as long as the job gets done.& Negative& too detailed\\
  \hline
        I found it really helpful, but made me slightly nervous thinking all my plans being successful are in the hands of ai tech& Mixed& helpful assistant, agency concerns \\
  \hline
  \textbf{Opinion towards Execution} &  \textbf{Sentiment}& \textbf{Reason}  \\
		\hline \hline
        The execution stage was amazing. I feel like this could be the future and we wont need to call or talk to people to get this kind of thing done ever again.& Positive& promising future\\
        \hline
        The execution stage went smoothly, except for a few rare instances of an error response before also saying the AI's automatic reply (which was correct).& Mixed& Smooth user experience, error response\\
		\hline
		I found it clunky and nit that user friendly& Negative& clunky, not user-friendly\\
        \hline
        This bit is user friendly, but very robotic, which makes it difficult to trust& Mixed& user-friendly, distrust due to robotic nature\\
		\hline
    \end{tabular}}
\end{table*}%

\subsubsection{Confidence Dynamics}
To visualize the user confidence in the planning and execution stage, we draw \revise{point plots} (see Figure~\ref{fig:confidence}) for user confidence in the task order. 
Overall, condition AP-AE shows the highest confidence in both the planning and execution stages. 
To verify the impact of user involvement in confidence, we adopted two-way ANOVA and post-hoc Tukey HSD test. 
We find that: (1) with user involvement in the planning, participants showed significantly lower confidence in planning (AP-AE > UP-AE, UP-UE); (2) with user involvement in the execution, participants showed a significantly lower confidence in execution (AP-AE > AP-UE, UP-UE). 
Meanwhile, users typically showed a higher confidence in the execution stage. 
Compared with conditions with automation execution (\ie condition AP-AE and UP-AE), the confidence gap narrows down in the conditions with user-involved execution (\ie condition AP-UE and UP-UE). 



\subsection{Analysis of Open Feedback} At the end of our study, we collected open feedback regarding the planning stage, execution stage, and any other feedback using the following question: 
`\textit{Please share any comments, remarks or suggestions regarding the planning/execution stage of LLM Assistant}' and `\textit{Do you have any other comments, remarks or suggestions regarding the study?}'. 
Overall, we analyzed all the feedback based on user opinions (positive, negative, mixed, neutral) and their suggestions. 
In our analysis, we ignored all phrases without any useful information like `None', `N/A', and `No comment'. 

\paratitle{\revise{Feedback and Suggestions}}. While most comments tended to show positive opinions (more than $80\%$) towards LLM agents as daily assistants, there are also negative opinions regarding the difficulty, expertise, trust, etc. We provide example excerpts from participants in Table~\ref{tab:example-opinion}. Besides opinions towards the system, some participants also appreciated our user-centric setup:"\textit{The study does a good job of emphasizing user experience by asking about perceptions of risk, trust, and confidence. This approach ensures that the evaluation is user-centric, which is important for assessing the real-world applicability of the LLM Assistant}."

Some participants also provided suggestions on how to further improve the design of LLM agent-based daily assistants. 
Regarding the plan edit, participants hope we can provide more convenient edit operations like `drop/drag' to adjust plan text ordering and `undo' operation to tolerate unexpected mistakes. 
Some participants also found the plans too detailed, which could increase the cognitive load (cf. Table~\ref{tab:example-opinion} except 3). 
As for the execution, many participants found it to be smooth. 
At the same time, they think additional verification in each step may further enhance the reliability of daily assistants: ``\textit{For the execution stage, I commend it for creating an input formatting box to execute the user's request validating each requirement}.''
There are also comments about the whole plan-then-execute workflow: ``\textit{The planning was really challenging, and I mostly left the default plans (they looked fine). 
This worked in the main, but a couple clearly needed revisiting. I would approach this iteratively: plan, test, observe, back to planning, then another test, before reaching the desired outcome}.'' \revise{Our findings suggest open research opportunities to explore more effective ways to provide an overview of plans that trade-off user cognitive load resulting from granular descriptions, with the need to provide details to help users identify flaws. For example, we can consider developing methods to interactively allow users to flesh out further details in a plan. }
\section{Discussion}
\label{sec:discussion}


\subsection{Key Findings}
\label{sec:discussion-findings}
Our experimental results show that user involvement in the plan-then-execute workflow with LLM agents can 
help fix imperfect plans in planning and wrong action predictions in the step-by-step execution. 
However, user involvement does not ensure a consistently positive impact on calibrated trust and overall task performance across different tasks. 

\paratitle{User Involvement Fails to Calibrate User Trust}. 
Overall, user involvement in the planning and execution does not significantly impact user trust and calibrated trust in planning and execution outcomes. 
As Table~\ref{tab:h1-res} shows, user involvement in planning can harm plan quality in tasks with a high-quality initial plan, which may potentially cause worse task performance in the subsequent execution stage. 
Our experimental results do not support \textbf{H1} or \textbf{H3}, which indicates user involvement does not necessarily help calibrate user trust in our study. 
Instead, with a task-specific correlation analysis (cf. Table~\ref{tab:correlation-2}), we found that the plan quality has a significant positive correlation with calibrated trust in both planning and execution outcomes. 
Combined with task-specific user trust and task-specific confidence, we can infer that users tend to trust the LLM agent overall. 
Such trust can be expected and calibrated in tasks with a high-quality plan. 
In contrast, users fail to calibrate their trust in the tasks where a low-quality plan is provided. {A potential cause of such miscalibrated trust is the plausibility of plans generated by LLMs (i.e., plans that appear to be likely correct).} 
In our study, all initial plans are formulated with a clear, logical structure, which covers most of the task requirements. 
At first glance, such high-quality text pieces seem quite plausible and trustworthy. 
We also received some open text feedback such as, --- ``The plans look nice, I do not find any space for improvement'' and ``the planning stage of the LLM assistant was helpful and trustworthy.'' 
Findings from recent work on LLM-assisted fact checking corroborate this, wherein authors found that convincing explanations provided by LLMs can cause over-reliance when LLMs are wrong~\cite{si2024large}.

\paratitle{User Involvement can Benefit Task Performance}. 
User involvement in planning and execution can positively impact overall task performance, especially execution accuracy. 
As the results in Table~\ref{tab:h1-res} and Table~\ref{tab:h2-res} show, user involvement in planning can help address imperfect plans (\eg task-1 with grammar error). Doing so further contributes to improvements in the execution accuracy. 
After controlling the plan quality, we found that user involvement in the execution can provide the best execution accuracy among most tasks considered in our study (cf. Table~\ref{tab:h4-res}). 
Based on the failure analysis (Section~\ref{sec-failure-analysis}), LLM agents can make mistakes in executing high-quality plans, which can be attributed to prediction errors (\ie wrong action name or action parameters) and prediction failures (\ie failure to provide valid action prediction). 
In practice with deployed LLM services, there is no reliability guarantee for the generated plan in planning or predicted actions in execution. 
User involvement can play an important role in the plan quality control and risky action control, ensuring \revise{that} only correct and {safe} actions \revise{are} executed to obtain desirable task outcomes. 

\paratitle{Other Findings}. 
We also found some user factors and perceptions that affect user trust and task performance. 
As seen in Table~\ref{tab:correlation}, nearly all covariates show a significant positive correlation with user trust in the AI system. 
Some of these covariates also impact user trust in the planning and execution outcomes. 
Overall, these findings indicate that users who are familiar with such systems tend to show higher user trust. 
However, some factors also correlate negatively with the calibrated trust in the execution outcomes and risk perception of using the LLM agents as daily assistants. 
This reflects that these factors can cause miscalibrated trust and reduced risk perception when working with the LLM agent. 
{While we found that risk perception negatively correlated with user trust, calibrated trust, task performance, and confidence (cf. Table~\ref{tab:correlation-2}), it does not mean risk perception is harmful in the human-LLM agent collaboration. 
The main cause is that users may only notice the risks of using LLM agents when the task is provided with a relatively low-quality plan.} 
Risk perception is important to calibrate user trust in the planning and execution outcomes. Collaborative workflows should support users with the provision to take over control of planning and/or execution stages based on their perceived risk.
 
\subsection{Implications}

\paratitle{The Impact of Convincingly Wrong LLM Outcomes}. 
As our study follows a plan-then-execute workflow for users to collaborate with LLM agents, users were not offered a chance to revise the plan after starting with execution. 
Users following a wrong plan can lead to negative outcomes. 
Combined with existing work on algorithm aversion~\cite{dietvorst2015algorithm} and the impact of negative first impressions on user trust~\cite{tolmeijer2021second}, we can infer that such convincingly wrong content~\cite{si2024large} can bias user trust and reliance towards the 
extremes. 
Before users take notice, they may develop an uncalibrated trust in the AI system, as observed through our findings in high-risk tasks (\ie tasks 1,2,3) and \revise{corroborating} work by~\citet{si2024large}. 
As a result, users over-rely on AI assistance, which is misuse akin to behavior that resonates with algorithm appreciation~\cite{logg2019algorithm}. 
Once users notice such phenomena, their trust in the LLM-based systems may sharply decrease, resulting in disuse due to algorithm aversion. 
{This can be a result of the misalignment between perceived AI performance and actual AI performance.
Existing human-AI collaboration literature has provided potential solutions for such problems, ranging from performance feedback interventions~\cite{he2023knowing} to agreement-in-confidence heuristic~\cite{Lu-CHI-2021,pescetelli2021role}.
} 
Future work can combine these insights to explore effective interventions for user trust calibration with convincingly wrong LLM outcomes.

\paratitle{Insights for Effective Collaboration with Plan-then-execute LLM Agents}. 
Our work has important theoretical implications for effective human-AI collaboration with plan-then-execute LLM agents. 
On the one hand, user involvement can be necessary to achieve complementary team performance. 
Although LLM agents have shown promising planning and execution capabilities, they are never perfect due to probabilistic uncertainty. 
With user involvement in the planning, users can fix imperfect plans with grammar errors (cf. Table~\ref{tab:h1-res} task-1). 
With user involvement in the execution, users can fix uncertainty issues (\eg LLM agent predicts invalid actions) and prevent risky actions (\eg LLM agents choose an itinerary conflicting with task requirements, cf. Table~\ref{tab:h4-res}, task-6). 
On the other hand, user involvement may also bring uncertainty and even harm LLM agent performance. 
In tasks where the LLM agent provides a high-quality plan (cf. task 4, 5, 6 in Table~\ref{tab:h1-res} and Table~\ref{tab:h2-res}), user involvement can harm the plan quality, which further negatively impacts the execution accuracy. 
Moreover, user involvement in planning and execution poses a significantly higher cognitive load on users (cf. Figure~\ref{fig:cognitive_load}) and negatively impacts user confidence (cf. Figure~\ref{fig:confidence}). 
Thus, too much human involvement in collaboration with plan-then-execute LLM agents can be undesirable. 
User involvement in the execution process brings more consistent benefits than user involvement in the planning stage. 
As suggested by the participants, iterative LLM agent simulation may be one potential way to decide when users should be involved. 
The LLM agent may first conduct a plan-then-execute round to obtain a clear plan and execution results. 
With humans checking the whole process and simulated outcomes, humans can decide whether to be involved in revising the plan or the execution process. 
In this way, we can minimize user involvement while keeping highly effective task outcomes through LLM agents.


\paratitle{\revise{Human Oversight and Designing Flexible Collaborative Workflows}}. 
\revise{In our study, we found that human oversight does not consistently lead to improved outcomes. 
One potential cause can be the disparity between the planning and execution of LLM agents. 
Specifically, it is unclear how one plan step will be transformed into one action. When users realize one plan step can be wrong during the execution stage, they may need to articulate it or manually override the agent action, posing a high cognitive load. 
Even worse, when users realize the LLM agent missed one action due to limited steps designed in the plan (in task-1), they do not have a chance to change the plan or add one extra step. 
To address such concerns, we may need a more flexible collaborative workflow where humans can fix planning and execution simultaneously.
In this way, users can exercise more flexible control over the workflow and the task outcomes. 
For instance, 
the action prediction from the LLM agent can be provided along with each step in the planning stage. 
Users can thereby be informed of the potential impact of their edited plan, which provides more straightforward feedback and helps users adjust the plan according to the expected actions.}

\subsection{Caveats and Limitations}
\label{sec-limitation}
\paratitle{Limitations and Potentail Biases}.
To ensure reliable task outcomes, humans are expected to fix imperfect plans (\eg grammar errors, misleading action intents) in the planning stage. 
However, not everyone in conditions UP-AE and UP-UE noticed such grammar errors and split the plan in task-1. 
Similarly, not everyone in conditions AP-UE and UP-UE noticed that the LLM agent chose an itinerary that conflicted with task requirements. 
As discussed earlier, LLM agents can generate plausible plans, which may mislead user trust in the planning and execution outcomes. 
In that case, participants in our study may have easily ignored some convincingly wrong plan steps or execution actions. 
In our study, one primary step in the plan is only transformed into a single action. In practice, LLM agents can generate multiple actions for one specific goal. 
However, such action generation and execution modes are challenging for humans to get involved in and control, as the execution of the action sequences is automated by the LLM agent within one goal. 
Furthermore, using multiple actions to achieve one primary step (\ie goal) also results in higher task complexity and reduced task clarity, which may impact the task outcomes 
~\cite{gadiraju2017clarity}. 

\paratitle{Transferability Concerns}. Although we selected representative tasks for daily scenarios, our study may not be enough to cover all potential cases of daily assistance with LLM agents. 
Some task characteristics (\eg task complexity, time consumption) may also impact how users are willing to rely on AI assistance. 
Meanwhile, full control over the plan-then-execute LLM agents may not be desirable for some simple tasks (\eg setting alarms). 
Once the efforts to control/interact with LLM agents are greater than the efforts to execute the tasks themselves, users will be unwilling to adopt such ``assistance." 
Future work can look into what daily user needs are suitable for LLM agents to support. 
\revise{In our study, the execution of plans is conducted in a simulation environment. While it has been proven to be effective in prior work of agent-based modeling and HCI studies~\cite{olson2014ways}, more work is needed to understand how execution of tasks in real-world environments with additional dependencies and complexities can influence our findings.} 

Participants in our study only followed a relatively fixed mode in collaboration with LLM agents, they can determine when to get involved in the planning and execution stages. 
The experimental conditions considered in our study range from full automation (\ie AP-AE) to full user control (\ie UP-UE). 
Such a setup provides good flexibility, which simulates real-world practice. 
Our findings and implications provide valuable insights to guide future research on human-AI collaboration with LLM agents.

\section{Conclusion}
This work empirically studied human-AI collaboration based on plan-then-execute LLM agents. 
Adopting such LLM agents in various everyday scenarios, 
we analyzed the impact of user involvement in the planning and execution stages on user trust and overall task performance. 
We provide various interactions in each stage to help users fix imperfect plans and modify execution outcomes. 
Our results suggest that the LLM agents can provide plausible text plans to cover task requirements, which can be convincingly wrong. 
As a result, users develop uncalibrated trust in the planning and execution outcomes, and user involvement in the planning and execution stages fails to calibrate user trust (\textbf{RQ1}). 
We also found that the plan quality substantially affects the subsequent execution accuracy. 
Thus, when user involvement in planning can fix imperfect plans, the overall task performance (\ie plan quality, accuracy of action sequence, and execution accuracy) gets improved. 
However, user involvement in planning can also harm task plan quality where the original plan is good to begin with. As a result, the LLM agents demonstrate worse task performance in these tasks. 
In contrast, 
user involvement in execution brings a more stable positive impact on task performance (\textbf{RQ2}). 
Our results suggest that plausible but wrong LLM outcomes can be detrimental to user trust calibration and overall task performance. 
We discussed the impact of convincingly wrong LLM outcomes and provided potential solutions and insights for future work. 
Furthermore, we synthesized key insights for better control and effective collaboration with plan-then-execute LLM agents. \revise{We also shed light on opportunities to design flexible collaborative workflows with human oversight for effective collaboration with LLM agents.} 

Our results indicate that user involvement in the LLM agent workflow can be important in ensuring reliable task outcomes. 
Future work can further investigate how to detect and handle plausible but imperfect LLM outcomes and design effective interventions to fix such problems. 
We hope that our key findings and implications reported in this work will inspire further research on human-AI collaboration with LLM agents.

\begin{acks}
This work was partially supported by the Delft Design@Scale AI Lab, the 4TU.CEE UNCAGE project, the Convergence Flagship ``ProtectMe'' project, by the Australian Research Council (ARC) Training Centre for Information Resilience (Grant No. IC200100022), and by an ARC Future Fellowship Project (Grant No. FT240100022). We made use of the Dutch national e-infrastructure with the support of the SURF Cooperative using grant no. EINF-5571 and EINF-9738. We finally thank all participants from Prolific and experts from our department.
\end{acks}

\bibliographystyle{ACM-Reference-Format}
\bibliography{LLM_Agent}

\clearpage
\appendix

\section{Details about Tasks}

Here, we provide more details about the selected task used in our study. For each task, we provide the automatic planning (\ie stepwise text plan) and execution (\ie action sequence) generated with our plan-then-execute LLM agent.

\paratitle{Task-1}: My account ID is 54321, and the password is PWD2023. I plan to make two foreign exchange transactions. The first is to buy 10,000 euros (with USD), and the second is to sell 5,000 US dollars (to EUR). Please help me operate.

\paratitle{Automatic Plan}:
\begin{itemize}
    \item 1. Log in to user account
    \begin{itemize}
        \item 1.1 Obtain account login information (Account ID: 54321, Account Password: PWD2023)
        \item 1.2 Log in to the account
        \item 1.3 Confirm successful login
    \end{itemize}
    \item 2. Conduct foreign exchange transactions
    \begin{itemize}
        \item 2.1 Buy euros
        \begin{itemize}
            \item 2.1.1 Obtain information for buying euros (Currency Type: EUR, Purchase Amount: 10000)
            \item 2.1.2 Buy the specified amount of euros
            \item 2.1.3 Confirm successful euro purchase
        \end{itemize}
        \item 2.2 Sell US dollars
        \begin{itemize}
            \item 2.2.1 Obtain information for selling US dollars (Currency Type: USD, Sell Amount: 5000)
            \item 2.2.2 Check the US dollar holdings
            \item 2.2.3 Obtain US dollar holdings information (Foreign Exchange Holdings Information: Obtained US dollar holdings information)
            \item 2.2.4 Sell the specified amount of US dollars
            \item 2.2.5 Confirm successful US dollar sale
        \end{itemize}
    \end{itemize}
\end{itemize}

\paratitle{Automatic execution}:
\begin{itemize}
    \item bank\_account\_login(account=54321, passwd=PWD2023)
    \item buy\_currency(account=54321, passwd=PWD2023, sourceCurrency=USD, targetCurrency=EUR, amount=10000)
\end{itemize}

\paratitle{Summary}: The plan is imperfect. Users are supposed to split the primary step 2 into two primary steps. Then, the LLM agent can conduct two transactions instead of one.

\paratitle{Task-2}: Please inquire about the current debt amount of my credit card with the last five digits 12345, and deduct the corresponding 12000 USD from my savings card number 6212345678900011 to repay this debt, then help me check the amount of the outstanding bill for the same credit card within 30 days after today.

\paratitle{Automatic Plan}:
\begin{itemize}
    \item 1. Inquire about the credit card debt amount
    \begin{itemize}
        \item 1.1 Obtain credit card information (Account ID: Account ID, Account Password: Account Password, Credit Card: Last five digits 12345, Verification Code: Verification Code)
        \item 1.2 Log in to the banking service platform
        \item 1.3 Inquire about the current credit card debt amount
        \item 1.4 Confirm the debt amount inquiry is successful and obtain the amount (Debt Amount: Inquired Debt Amount)
    \end{itemize}
    \item 2. Deduct funds to repay the credit card debt
    \begin{itemize}
        \item 2.1 Obtain savings card information (Savings Card Number: 6212345678900011)
        \item 2.2 Transfer from savings card to credit card
        \item 2.3 Confirm the transfer is successful
    \end{itemize}
    \item 3. Check the credit card's future outstanding bills
    \begin{itemize}
        \item 3.1 Obtain date information (Date Range: Within 30 days after today)
        \item 3.2 Inquire about the credit card's outstanding bill amount within the specified date range
        \item 3.3 Confirm the future outstanding bill amount inquiry is successful and obtain the amount (Outstanding Bill Amount: Inquired Outstanding Bill Amount)
    \end{itemize}
\end{itemize}

\paratitle{Automatic execution}:
\begin{itemize}
    \item check\_balance(card=12345)
    \item pay\_credit\_card(source\_card=6212345678900011, target\_card =12345, amount=12000)
    \item check\_credit\_card\_bills(card=12345, period=30)
\end{itemize}

\paratitle{Summary}: The authors need to revise the plan and add one primary step to search for the credit card, which ends with `12345'. Due to missing this step, the LLM agent can not correctly execute all tasks because the correct credit card ID (\ie 4980981092312345) is missing.

\paratitle{Task-3}: I need to schedule a repair for my TV at 6 PM tomorrow evening. The brand is Sony, model X800H, and there is an issue with the screen. Please book the repair service and tell me the
reservation number.

\paratitle{Automatic Plan}:
\begin{itemize}
    \item 1. Obtain user contact information
    \begin{itemize}
        \item 1.1 Obtain user's name (Name: User's name)
        \item 1.2 Obtain user's contact number (Contact number: User's phone number)
        \item 1.3 Obtain the installation address of the user's TV (Installation address: User's address)
    \end{itemize}
    \item 2. Find repair service
    \begin{itemize}
        \item 2.1 Determine the repair service provider
        \begin{itemize}
            \item 2.1.1 Search for authorized Sony TV repair service providers
            \item 2.1.2 Select the nearest service provider to the user (Service provider list: Authorized Sony TV repair service provider list, User address: User's address)
        \end{itemize}
        \item 2.2 Obtain the contact information of the repair service provider (Service provider contact information: Nearest service provider's contact information to the user)
    \end{itemize}
    \item 3. Book repair service
    \begin{itemize}
        \item 3.1 Contact the repair service provider
        \begin{itemize}
            \item 3.1.1 Provide the brand, model, and issue description of the TV (Brand: Sony, Model: X800H, Issue description: Screen issue)
            \item 3.1.2 Provide the user's contact information (Name: User's name, Contact number: User's phone number, Installation address: User's address)
            \item 3.1.3 Obtain the appointment time (Appointment time: Tomorrow's date 18:00:00)
            \item 3.1.4 Book the repair service
        \end{itemize}
        \item 3.2 Obtain the reservation number (Reservation status: Repair request status, Reservation number: Repair request number)
    \end{itemize}
    \item 4. Notify the user
    \begin{itemize}
        \item 4.1 Send the reservation number and the contact information of the repair service provider to the user (Reservation number: Repair request number, Service provider contact information: Repair service provider's contact information)
    \end{itemize}
\end{itemize}

\paratitle{Automatic execution}:
\begin{itemize}
    \item obtain\_user\_info()
    \item search\_service\_provider(brand=Sony)
    \item appliance\_repair\_request(applianceModel=X800H, applianceType=TV, appointment\_time=Tomorrow 6 PM,issueDescription =Screen issue, servicepointID=[Sony01, Sony03])
    \item appliance\_repair\_request(applianceModel=X800H, applianceType=TV, appointment\_time=Tomorrow 6 PM,issueDescription =Screen issue, servicepointID=[Sony01, Sony03])
\end{itemize}

\paratitle{Summary}: While the first three primary steps are clear, the last primary step is not aligned with any provided actions. Users are supposed to delete the last primary step or override the potentially unnecessary actions after the first three steps.

\paratitle{Task-4}: I need to set an alarm for every weekday morning at 7:30, and then cancel the alarm for Thursday, changing it to 8:00 in the evening.

\paratitle{Automatic Plan}:
\begin{itemize}
    \item 1. Set the alarm for every day
    \begin{itemize}
        \item 1.1 Get the alarm setting information (Time: 07:30 AM, Frequency: Monday to Friday)
        \item 1.2 Set the alarm
        \item 1.3 Confirm whether the alarm is set successfully
    \end{itemize}
    \item 2. Cancel the Thursday alarm
    \begin{itemize}
        \item 2.1 Get the information of the alarm to be cancelled (Time: 07:30 AM, Frequency: Thursday)
        \item 2.2 Cancel the alarm
        \item 2.3 Confirm whether the alarm is cancelled successfully
    \end{itemize}
    \item 3. Set the new alarm time
    \begin{itemize}
        \item 3.1 Get the information of the alarm to be adjusted (New Time: 08:00 PM, Frequency: Thursday)
        \item 3.2 Set the new alarm
        \item 3.3 Confirm whether the new alarm is set successfully
    \end{itemize}
\end{itemize}

\paratitle{Automatic execution}:
\begin{itemize}
    \item create\_alarm(hour=7, minute=30, repeat=Weekly, frequency =Weekday)
    \item cancel\_alarm(hour=7, minute=30, repeat=Weekly, frequency =Thursday)
    \item create\_alarm(hour=20, minute=0, repeat=Weekly, frequency =Thursday)
\end{itemize}

\paratitle{Summary}: Both the automatic planning and execution are correct.

\paratitle{Task-5}: I have an important meeting to attend next Wednesday, and I need to book a flight ticket from London to Amsterdam for tomorrow, it must be a morning flight, and then return from
Amsterdam to London tomorrow night, please handle it for me.

\paratitle{Automatic Plan}:
\begin{itemize}
    \item 1. Search outbound flight ticket
    \begin{itemize}
        \item 1.1 Obtain outbound flight search information (Departure City: London, Arrival City: Amsterdam, Departure Date: Tomorrow's Date, Departure Time Slot: Morning)
        \item 1.2 Search for outbound flights
        \item 1.3 Select a suitable outbound flight (Flight Number: Selected Flight Number)
    \end{itemize}
    \item 2. Book outbound flight ticket
    \begin{itemize}
        \item 2.1 Obtain outbound passenger information (Name: User Name, ID Number: User ID Number)
        \item 2.2 Book outbound flight ticket
        \item 2.3 Confirm whether the outbound flight ticket is successfully booked
    \end{itemize}
    \item 3. Search return flight ticket
    \begin{itemize}
        \item 3.1 Obtain return flight search information (Departure City: Amsterdam, Arrival City: London, Departure Date: Tomorrow's Date, Departure Time Slot: Evening)
        \item 3.2 Search for return flights
        \item 3.3 Select a suitable return flight (Flight Number: Selected Flight Number)
    \end{itemize}
    \item 4. Book return flight ticket
    \begin{itemize}
        \item 4.1 Obtain return passenger information (Name: User Name, ID Number: User ID Number)
        \item 4.2 Book return flight ticket
        \item 4.3 Confirm whether the return flight ticket is successfully booked
    \end{itemize}
\end{itemize}

\paratitle{Automatic execution}:
\begin{itemize}
    \item search\_flight(departure\_city=London, arrival\_city=Amsterdam, date=Tomorrow, timePeriod='Morning')
    \item book\_flight(flight\_id=flight\_id, flight\_class=economics)
    \item search\_flight(departure\_city=Amsterdam, arrival\_city=London, date=Tomorrow, timePeriod='Evening')
    \item book\_flight(flight\_id=flight\_id, flight\_class=economics)
\end{itemize}

\paratitle{Summary}: Both the automatic planning and execution are correct.

\paratitle{Task-6}: Please plan a trip for me departing on October 1st at 8:00 AM to Japan, returning on October 7th at 11:00 PM, including Tokyo Disneyland, Senso-ji Temple, Ginza, Mount Fuji, Kyoto cultural experience, Universal Studios Osaka, and visiting the Nara Deer Park on October 4th, and help me find hotels where the nightly cost does not exceed 10,000 Japanese yen.

\paratitle{Automatic Plan}:
\begin{itemize}
    \item 1. Itinerary planning
    \begin{itemize}
        \item 1.1 Obtain itinerary information (destination: Japan, starting from October 1st 8:00 AM, return around October 7th 11:00 PM, including Tokyo Disneyland, Senso-ji Temple, Ginza, Mount Fuji, Kyoto cultural experience, Universal Studios Osaka, and visiting the Nara Deer Park on October 4th)
        \item 1.2 Plan the itinerary according to the info
        \item 1.3 show all planned itineraries
    \end{itemize}
    \item 2. Itinerary Selection
    \begin{itemize}
        \item 2.1 Show users the planned itinerary
        \item 2.2 ask users to select one route from them
        \item 2.3 confirm the selected itinerary and show the schedule
    \end{itemize}
    \item 3. Book hotels
    \begin{itemize}
        \item 3.1 Obtain travel budget (Accommodation budget: Not exceeding 10,000 yen per night)
        \item 3.2 Search for hotels within budget
        \item 3.3 arrange hotels for each night within budget (nightly cost does not exceed 10,000 Japanese yen)
        \item 3.4 Complete hotel booking and show the arranged hotel plan
    \end{itemize}
\end{itemize}

\paratitle{Automatic execution}:
\begin{itemize}
    \item travel\_itinerary\_planner(destination=Japan, departure\_time= October 1st 8:00 AM, return\_time=October 7th 11:00 PM, interests=Tokyo Disneyland, Senso-ji Temple, Ginza, Mount Fuji, Kyoto cultural experience, Universal Studios Osaka, and visiting the Nara Deer Park on October 4th)
    \item select\_itinerary(selection=route-1)
    \item book\_hotel(selection=route-1)
\end{itemize}

\paratitle{Summary}: The plan is correct. However, the LLM agent makes mistakes in selecting the route. The expected route should be route-2 instead of route-1. The wrong parameters will further affect the hotel booking in step 3.

\end{document}